\documentclass[prb,twocolumn,amsmath,a4paper,superscriptaddress]{revtex4}

\usepackage{graphicx}

\begin{document}
\title{Superconducting instability in 3 bands metallic nanotubes}

\author{David Carpentier}
\affiliation{CNRS  UMR 5672 - Laboratoire de Physique de l'Ecole Normale
Sup{\'e}rieure de Lyon, \\
46, All{\'e}e d'Italie, 69007 Lyon, France}

\author{Edmond Orignac}
\affiliation{CNRS UMR 5672 - Laboratoire de Physique de l'Ecole Normale
Sup{\'e}rieure de Lyon, \\
46, All{\'e}e d'Italie, 69007 Lyon, France}

\date{\today}

\begin{abstract}
  Motivated by recent experiments on small radius nanotubes, we
study the superconducting instabilities of cylindrical (5,0)
nanotubes. According to  band structure calculations,  these
nanotubes 
possess three bands at the Fermi energy. 
 Using a fermionic renormalization group approach and a careful 
bosonization treatment,
we consider  the effect of
different attractive interactions, mediated by phonons, within the 
Luttinger Liquid framework. We particularly focus on a superconducting 
instability specific to
the three bands model we consider for the description of  these
 (5,0) cylindrical nanotubes. 
\end{abstract}

\maketitle

\section{Introduction}
\label{sec:intro}

Superconducting behavior of Carbon nanotubes have been observed
 both in ropes of single walled nanotubes\cite{Kociak:2001}, and in
  single walled\cite{Tang:2001} and
 multiwalled\cite{Takesue:2005} small radius nanotubes grown  in a
 zeolite matrix.  The high value of the critical temperature $T_{c}$ in these
 last two cases raised the question of its possible relation with the
 small radius of the nanotubes. Indeed, for this 4 angstr{\"o}m
 nanotubes, the large curvature induces an hybridization of the
 $\sigma$ and $\pi$ orbitals of the carbon atoms\cite{Blase:1994},
 leading to electron and phonon properties different from the larger
 nanotubes. The relation between the high $T_{c}$ and these
 peculiarities have motivated several works, in particular on the
 metallic (5,0) nanotube which constitute the best candidate for the
 origin of the superconductivity\cite{Sedeki:2002}.  These previous
 approaches include 
 both numerical calculations of the band structure and phonon dispersion
 relation\cite{Connetable:2005,Barnett:2005}, and  renormalization group
 approaches either restricted 
to a subspace of the couplings\cite{Kamide:2003}, or
 using specific initial conditions in the full space 
of couplings.\cite{Gonzalez:2005}

 In this paper, we identify the different instabilities of the (5,0)
 metallic nanotubes in the presence of effective electronic attractive
 couplings mediated by phonons.
We follow previous approaches on
 larger nanotubes\cite{Egger:1997,egger_nanotubes,balents97_nanotubes,kane97_nanotubes,yoshioka99_nanotubes,odintsov99_nanotubes,odintsov_nanotubes,lin98_nanotubes,Sedeki:2002,levitov_nanotube,nersesyan03_nanotube} 
in using the Luttinger Liquid
 framework to describe the low energy behavior of nanotubes.
 Our approach is based on the
band structure for {\it cylindrical } (5,0) nanotubes provided by
 various methods such as the
Local Density Approximation (LDA), the GW method,  
and tight binding calculations, consisting in three bands at the Fermi
 energy.\cite{Blase:1994,li01_small_radius_nanotubes,machon02_nanotube_bands,liu02_nanotube_bands,miyake03_nanotubes_gw} 
 Then, we study the perturbations of this band structure induced by
the residual interactions between the low energy fermions.
The nature of these interactions is constrained by the
specific symmetries of the initial band structure, different from
usual 3-leg fermionic ladders previously studied in
 Refs.~\onlinecite{arrigoni_3chain,kimura96_3chain,schulz_moriond,lin97_nchains,rice97_3chain,white98_3chains_dmrg,kimura98_3chain,yonemitsu99_3chains_dmrg,ledermann00_nchains,tsuchiizu01_3chains}. Using a
fermionic renormalization group we identify the dominant instability
corresponding to each effective attractive potential. The
 instabilities we find involve either electronic degrees of freedom
 on a single band, or on a two band subsystem, or on the whole three
 band system. We focus on this last case, which corresponds to
 new instabilities specific to the symmetries of the (5,0) nanotubes.
This allows to
 determine the momentum of the phonons responsible for the main
 instability, analogously to the proposal for a superconductivity
 induced by radial breather modes in regular 2-bands metallic
 nanotubes\cite{deMartino:2004}. 
 For
 the specific instabilities, seen as strong coupling directions of the
 renormalization group flow, we use the abelian bosonization
 formalism to identify its nature and specify the corresponding
 dominant correlation function. This bosonization description
 requires a careful treatment of the so-called Klein factors, a
 crucial technical point in this 3 bands model. 
  We pedagogically present this problem and its solution in section
 \ref{sec:phases}. Finally, the remaining gapless spin modes are
 identified using a non-abelian bosonization
 approach.\cite{novikov82_wz,witten_wz,polyakov83_wz,knizhnik_wz,affleck_wz,affleck_strongcoupl}   
 
  The paper is organized as follows : in section \ref{sec:model} we
  define the fermionic  model we consider and the notations used
  throughout the paper. The renormalization approach is sketched in the
  next section \ref{sec:RG-fermions}, the details being postponed to
  appendix  \ref{app:RG-deriv} for readability of the manuscript. The
  strong coupling phases  are analyzed in sec. \ref{sec:phases}, as
  well as the conventions used for the bosonization formalism. The
  complete bosonized expressions of all operators and necessary correlations
  functions are given in appendix \ref{app:bosonization}. Finally we
  discuss the validity of our results and our main conclusions in
  section \ref{sec:conclusion}. The appendix \ref{app:klein} is
  devoted to some peculiar technical difficulties of our model
  associated with the Klein factor of abelian bosonization.

\section{Model}
\label{sec:model}

\subsection{Band structure of (5,0) nanotube}\label{sec:bandes}

\begin{figure}[htbp]
\centerline{\includegraphics[width=7cm]{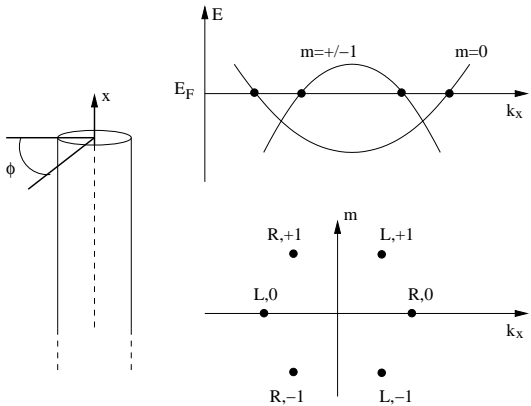}}
\caption{Schematic representation of the band structure near $E_{F}$
for the  3-bands nanotubes considered in this paper.}
\label{fig:bandes}
\end{figure}

 The band structure predicted by LDA-DFT calculations for (5,0)
metallic nanotubes is depicted schematically around the Fermi energy
$E_{F}$ in Fig.\ref{fig:bandes}. It consists of 
three
bands\cite{Blase:1994,li01_small_radius_nanotubes,liu02_nanotube_bands,machon02_nanotube_bands,miyake03_nanotubes_gw}. 
For a cylindrical nanotube, rotational invariance
results in the conservation of the angular momentum $m$, and
translational invariance 
in the conservation of momentum along the tube $k_{x}$. 
The quantum numbers of the three bands are thus determined
accordingly. In our specific case  
the two bands with angular momentum $m=\pm 1$ are degenerate and 
possess the same Fermi momentum $k_{F_{1}}$, smaller that the 
Fermi momentum $k_{F_{0}}$ of the band $m=0$. 
Linearizing these bands near $E_{F}$, we decompose the fermion
annihilation operator into :
\begin{equation}\label{eq:def-psi}
\psi_{0,\sigma}(x,\phi) = e^{i k_{F_{0}} x } \psi_{R,0,\sigma}(x) \
  + e^{-i k_{F_{0}}x } \psi_{L,0,\sigma}(x)
\end{equation}
 and
\begin{multline}
\psi_{m=\pm 1,\sigma}(x,\phi) = e^{-i k_{F_{1}}x + i m \phi } \psi_{R,m,\sigma}(x) \\
  + e^{+i k_{F_{1}}x - i m \phi} \psi_{L,m,\sigma}(x)
\end{multline}
where  $\psi_{R/L,m,\sigma}$ represent the
annihilation operator for a right (resp. left) moving fermion of angular
momentum $m$ and spin $\sigma$. From now on $x$ denotes the
coordinate in the direction of the nanotube, and $\phi$  the angle along the
circumference (see Fig.~\ref{fig:bandes}).  Note that as the Fermi velocity
$v_{F_{1}}$ in  the $m=\pm 1$ bands in negative, the right moving
fermions of these bands have a longitudinal momentum $-k_{F_{1}}$ as
opposed to the usual situation of the band $m=0$ where they have
momentum $+k_{F_{0}}$( see Fig.\ref{fig:bandes}).  
We then describe the low energy
properties of this model by the simple Hamiltonian
\begin{equation}
H_{0}^{(0)} = - \sum_{\sigma=\uparrow,\downarrow}
\sum_{m=0,\pm 1} v_{F_{m}}
\int dx ~
\psi^{\dagger}_{R,m,\sigma} \partial_{x}  \psi_{R,m,\sigma}^{} . 
\end{equation}
 Note that as the nature of the superconducting instabilities we
 discuss in this paper will not depend on the differences of Fermi
 velocities between the bands, we will assume from now that they are
 all equals : $v_{F_{1}} = v_{F_{0}} = v_{F}$.

\subsection{Residual interactions}\label{sec:interactions}

Then we consider the perturbations around this band structure,  which
can originate either from electronic interactions not taken into
account by the band structure calculations, especially in a quasi-1D geometry,
or from the effective attractive interaction originating from the coupling
to phonons. Within our effective low-energy approach, we consider a
minimal model  possessing all the
conservation laws. This results in a perturbative interaction action
which can be written formally as 
\begin{multline}\label{eq:formal-Sint}
S_{\textrm{int}} = 
g^{(1)}_{abcd}\sum_{\sigma,\sigma'}\int dx~
\psi^{\dagger}_{R,a,\sigma} \psi^{\dagger}_{L,b,\sigma'}
\psi_{R,c,\sigma'}\psi_{L,d,\sigma }\\
+ g^{(2)}_{abcd}\sum_{\sigma,\sigma'}\int dx~
\psi^{\dagger}_{R,a,\sigma} \psi^{\dagger}_{L,b,\sigma'}
\psi_{L,c,\sigma'}\psi_{R,d,\sigma}
\end{multline}
 where $a,b,c,d = 0,\pm 1$ stands for the band indices (angular
momentum). In this expression, as is usual in 1D systems, the first
part correspond to the 
back scattering operators, and the second to forward scattering
operators. The forward scattering part can be
decomposed into $g_4$ processes and $g_2$
processes.\cite{solyom_revue_1d,schulz_houches_revue} 
The $g_4$ processes only renormalize
the velocities of the
particles.\cite{solyom_revue_1d,schulz_houches_revue}
 Since we have neglected the 
velocity differences, consistency requires to also neglect the $g_4$
processes.  
We use the convention for the action that the
partition function of the system is written as 
$Z=\int d\psi d\psi^{\dagger}\exp ( -S[\psi ,\psi^{\dagger}]$),
and that repulsive (resp. attractive)  interactions between the fermions
 correspond to $g^{(1)},g^{(2)}>0$ (resp. $<0$).

 To proceed, we must use the symmetry of the problem at sake to
select out of all the couplings in (\ref{eq:formal-Sint}) only those
fulfilling the required conservation laws. As the results of band
structure calculation
suggest, the Fermi wavevector $k_{F_{1}}$ of the two bands $\pm 1$ is 
different (and incommensurate) from the Fermi wavevector
$k_{F,0}\ne k_{F,1}$ of the band with angular momentum $m=0$. 
 Interactions must preserve both rotational invariance and
 translational invariance, {\it i.e}  conserve the total angular
 momentum $m$ and the total momentum $k_{x}$.
 To classify these interactions, we  follow the notations of
 Refs.~\onlinecite{Krotov:1997} and \onlinecite{Gonzalez:2005}. 
 Note that whereas this model has some superficial similarity with the
 3-leg ladder 
model\cite{arrigoni_3chain,kimura96_3chain,lin97_nchains,ledermann00_nchains,tsuchiizu01_3chains}, it differs from it by the symmetries as  all three Fermi
momentum
 are different, as opposed to the present case. 

\subsubsection{Interactions in the band $m=0$ subsystem.}

  The first two allowed interactions are the usual back-scattering
and forward scattering interactions in the single band $m=0$. The
associated fields are denoted by $g^{(1)}$ and $g^{(2)}$ and they
correspond to the action 
\begin{multline}\label{eq:Sint-0}
S_{\textrm{int}}^{(0)} =
-g^{(1)} \sum_{\sigma,\sigma'} \int dx~
\psi^{\dagger}_{R,0,\sigma}\psi^{\dagger}_{L,0,\sigma'}
\psi^{}_{L,0,\sigma}\psi^{}_{R,0,\sigma'}
\\
+ g^{(2)} \sum_{\sigma,\sigma'} \int dx~
\psi^{\dagger}_{R,0,\sigma}\psi^{\dagger}_{L,0,\sigma'}
\psi^{}_{L,0,\sigma'}\psi^{}_{R,0,\sigma}
\end{multline}

\subsubsection{Interactions in the two bands $m=\pm 1$ subsystem.}

\begin{figure}[tbp]
\centerline{\includegraphics[height=5cm]{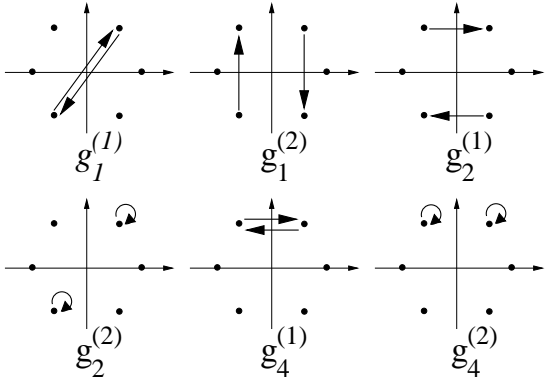}}
\caption{Formal representation in the $k_{x},m$ plane of the
 considered interactions in the two
band $m=\pm 1$ subsystem. }
\label{fig:operateur-2bandes}
\end{figure}

The next group of interactions we consider are the forward and
back-scattering couplings in the subsystem consisting in the two
degenerate bands $m=\pm 1$. This corresponds exactly to a two leg
ladder with degenerate bands.\cite{varma_2chain,penc_2chain,fabrizio_2ch_rg,kuroki94_2ch,khveshenko_2chain,nagaosa_2ch,schulz_2chains,nagaosa_chiral_anomaly_1d,balents_2ch,yoshioka_2ch,shelton_tj_ladder,schulz_moriond,lee99_2ch,abramovici05_2ch} Note that even this 2 band subsystem
differs from the usual description of larger nanotubes, which
possess right (or left) moving fermions at both $+k_{F}$ and
$-k_{F}$.\cite{ando05_nanotube_review} 
The interactions are depicted
 schematically in Figure \ref{fig:operateur-2bandes}. The explicit
 part of the interacting action  is

\begin{align}
& S_{\textrm{int}}^{(\pm1)}  = \sum_{\sigma,\sigma'} \int dx~ \\ \nonumber 
& - g_{1}^{(1)} \left(
\psi^{\dagger}_{R,+1,\sigma}\psi^{\dagger}_{L,-1,\sigma'}
\psi^{}_{L,-1,\sigma}\psi^{}_{R,+1,\sigma'} + (+1\leftrightarrow -1)
\right) \\ \nonumber 
&+g_{1}^{(2)} \left(
\psi^{\dagger}_{R,+1,\sigma}\psi^{\dagger}_{L,-1,\sigma'}
\psi^{}_{L,+1,\sigma'}\psi^{}_{R,-1,\sigma}
 + (+1\leftrightarrow -1)
\right) \\ \nonumber 
&- g_{2}^{(1)} \left(
\psi^{\dagger}_{R,+1,\sigma}\psi^{\dagger}_{L,-1,\sigma'}
\psi^{}_{L,+1,\sigma}\psi^{}_{R,-1,\sigma'} 
+ (-1\leftrightarrow -1)
\right) \\ \nonumber
&+ g_{2}^{(2)} \left(
\psi^{\dagger}_{R,+1,\sigma}\psi^{\dagger}_{L,-1,\sigma'}
\psi^{}_{L,-1,\sigma'}\psi^{}_{R,+1,\sigma}
 + (+1\leftrightarrow -1)
\right) \\ \nonumber
&- g_{4}^{(1)} \left(
\psi^{\dagger}_{R,+1,\sigma}\psi^{\dagger}_{L,+1,\sigma'}
\psi^{}_{L,+1,\sigma}\psi^{}_{R,+1,\sigma'} 
+ (+1\leftrightarrow -1)
\right)\\ \nonumber
&+ g_{4}^{(2)} \left(
\psi^{\dagger}_{R,+1,\sigma}\psi^{\dagger}_{L,+1,\sigma'}
\psi^{}_{L,+1,\sigma'}\psi^{}_{R,+1,\sigma}
+ (+1\leftrightarrow -1)
\right)
\end{align}

\subsubsection{Interactions between the $m=0$ band, and the two bands
$m=\pm 1$} 

\begin{figure}[htbp]
\centerline{\includegraphics[height=5cm]{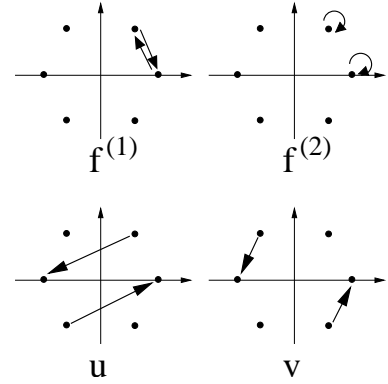}}
\caption{Formal representation in the $k_{x},m$ plane of the
 considered interactions between the $m=0$ band and the two
band $m=\pm 1$ subsystem. }
\label{fig:operateur-3bandes}
\end{figure}

 The last group of interactions, specific to the model we consider,
 corresponds to the interactions between the band $m=0$ and the two band
 subsystem $m=\pm 1$. With our conventions for the signs of the
 coupling, they read
\begin{align}
& S_{\textrm{int}}^{(0/\pm1)}  = \sum_{\sigma,\sigma'} \int dx~ \\
\nonumber
&- f^{(1)} \left(
\psi^{\dagger}_{R,0,\sigma}\psi^{\dagger}_{L,+1,\sigma'}
\psi^{}_{L,+1,\sigma}\psi^{}_{R,0,\sigma'} 
+ (+1\leftrightarrow -1)+h.c.
\right)\\ \nonumber 
&+ f^{(2)} \left(
\psi^{\dagger}_{R,0,\sigma}\psi^{\dagger}_{L,+1,\sigma'}
\psi^{}_{L,+1,\sigma'}\psi^{}_{R,0,\sigma}
+ (+1\leftrightarrow -1)+h.c.
\right)\\ \nonumber 
&+ u\left(
\psi^{\dagger}_{R,0,\sigma}\psi^{\dagger}_{L,0,\sigma'}
\psi^{}_{L,+1,\sigma'}\psi^{}_{R,-1,\sigma}
+ (+1\leftrightarrow -1)+h.c.
\right)\\ \nonumber 
&-v \left(
\psi^{\dagger}_{R,+1,\sigma}\psi^{\dagger}_{L,-1,\sigma'}
\psi^{}_{L,0,\sigma}\psi^{}_{R,0,\sigma'} 
+ (+1\leftrightarrow -1)+h.c.
\right)
\end{align}
 All these couplings are depicted schematically in
 Fig.~\ref{fig:operateur-3bandes}. 

\section{Renormalization group study}
\label{sec:RG-fermions}

\subsection{Derivations of the scaling equations}

 Having defined explicitly the action describing our model, we will
now study its low energy behavior using the renormalization  group
formalism. The standard procedure for one-dimensional fermionic model
is implemented by using the operator product expansion formalism (see
{\it e.g} Ref.~\onlinecite{cardy_scaling}, chap.5).  The product expansion of the
four fermions operators  
appearing in the perturbative expansion of the partition function
reads formally 
\begin{align} \label{eq:OPE}
 \langle 
\psi^{\dagger}_{R,a} \psi^{\dagger}_{L,b}   \psi_{L,c} & \psi_{R,d}
\psi^{\dagger}_{R,e} \psi^{\dagger}_{L,f} \psi_{L,g} \psi_{R,h} 
\rangle =  \\ \nonumber 
&+\frac{\delta_{ah}\delta_{bg}}{4\pi^{2} z_{a}\overline{z}_{b} }
\langle \psi^{\dagger}_{R,e} \psi^{\dagger}_{L,f} 
        \psi_{L,c} \psi_{R,d}\rangle \\ \nonumber 
&-\frac{\delta_{ah}\delta_{cf}}{4\pi^{2} z_{a}\overline{z}_{c} }
\langle \psi^{\dagger}_{R,e} \psi^{\dagger}_{L,b}
        \psi_{L,g} \psi_{R,d}\rangle \\ \nonumber 
&-\frac{\delta_{bg}\delta_{de}}{4\pi^{2} z_{d}\overline{z}_{b} }
\langle \psi^{\dagger}_{R,a} \psi^{\dagger}_{L,f}
        \psi_{L,c} \psi_{R,h}\rangle \\ \nonumber 
&+\frac{\delta_{cf}\delta_{de}}{4\pi^{2} z_{d}\overline{z}_{c} }
\langle \psi^{\dagger}_{R,a} \psi^{\dagger}_{L,b}
        \psi_{L,g} \psi_{R,h}\rangle
\end{align}
where we have used mixed labels $a,b,c,d,e,f,g,h$ for the band $m$ and spin
$\sigma$.   In this expression, $z_{a}$ stands for
$x-iv_{a}\tau$. Within the approximation $v_{F_{a}}=v_{F}$, all the
prefactor will produce the constant  \begin{equation}
\frac{1}{4\pi^2 v_{F} } \int_{a < |z| < a e^{dl}}  \frac{ dz
d\overline{z} }{z\overline{z}} =  
\frac{dl }{2\pi v_{F}}  , 
\end{equation}
\noindent where $a$ is a real space ultraviolet cutoff. 
 Thus, specifying the operator product expansion (\ref{eq:OPE}) to
the interactions 
of our model, we obtain the renormalization group equations to second order 
in the couplings $g_{i}^{(j)},f_{i},u,v$. They are given explicitly
in  formula 
(\ref{eq:RGeqs}) in  appendix \ref{app:RG-deriv}. They only  differ from
those of Ref.~\onlinecite{Gonzalez:2005} by an extra term $2 (u^{2}
+v^{2})$ in 
the equation for $\partial_{l}g_{2}^{(2)}$. 

\subsection{Renormalization flow integration}

The scaling equations (\ref{eq:RGeqs}) admit asymptotic solutions of
the form 
\begin{equation}\label{eq:ansatz}
g_{i}^{(j)} (l) = \frac{c_{ij}}{(l^{*}-l)^{\mu_{ij}}} + \mathcal{O}
((l-l^{*})^{-\mu_{ij}}). 
\end{equation}
However, a direct analytical solution in the full parameter space is
not tractable.  Hence we have numerically integrated these equations
for different initial values of the couplings. Our strategy was to
choose some reasonable perturbative initial point in the parameter
space, and to study the instability from this starting point 
occuring upon increasing the
strength of a given attractive interaction. We have done this for all
the possible attractive interactions, {\it i.e} the different phonons
coupling the different electronic branches of the model : 
$g^{(1)},f^{(1)},g_{1}^{(1)},g_{1}^{(2)},g_{2}^{(1)},g_{4}^{(1)},u,v$.
We have also checked that the results were independent of the initial
perturbative point chosen.

 Each of these instabilities corresponds to a strong coupling
direction  where at least some of the couplings $g_{i}^{(j)},f^{(i)},u,v$
diverge at a finite scaling length $l^{*}$. Thus we characterize each
strong coupling direction by the subset of the most diverging
couplings, namely those with the largest power $\mu_{ij}$ in
Eq.~(\ref{eq:ansatz}).  Indeed, 
we have found that for the considered directions, while the 
dominant couplings always diverge with an exponent $\mu_{ij}=1$, 
there exist other couplings diverging with smaller exponents
$\mu_{ij}<1$. These couplings with weaker divergences are expected to
give rise to anomalous scaling\cite{lin00_scaling,konik02_symmetry}
but  not to modify the strong coupling phases.

The nature of the instability corresponding to a given strong
coupling fixed point will be identified in the
section~\ref{sec:phases}, by
bosonizing the model in the subspace consisting of the dominant
diverging couplings. We will focus particularly on the
instabilities specific to the three band model.

\subsubsection{Single band or two-band model
  instabilities}\label{sec:single-band} 

  We first list the instabilities of the band $m=0$, and two bands $m=\pm 1$ 
subsystem. These instabilities are not specific to the present model, and have 
been previously studied (see
e.g. Refs.~\onlinecite{giamarchi_book_1d,gogolin_book} and references therein). 
The first instability is obtained
for a negative  $g_{1}$, {\it i.e}  an attractive interaction in the $m=0$
band. The  
corresponding asymptotic fixed point, at which $g_{1}$ and $g_{2}$ 
both diverge to $-\infty$, is the well known instability of the Luther-Emery 
model.\cite{luther_exact,solyom_revue_1d,emery_revue_1d}  

 Three different instabilities affect only the bands $m=\pm 1$. 
Upon decreasing $g_{1}^{(1)}$ to negative values, we find that the
dominant diverging couplings are 
$g_{1}^{(1)},g_{1}^{(2)},g_{2}^{(1)},g_{2}^{(2)}$ and $g_{4}^{(2)}$. While 
$g_{1}^{(1)}$ and $g_{2}^{(2)}$ flow towards $-\infty$, $g_{1}^{(2)},g_{2}^{(1)}$
and $g_{4}^{(2)}$ flow towards $+\infty$. 
   The phase associated with a negative coupling  $g_{1}^{(2)}<0$ is described 
 by : dominant divergence of $g_{2}^{(1)} \to + \infty$ and 
 $g_{2}^{(2)},g_{4}^{(1)},g_{4}^{(2)}\to - \infty$. 
 Finally a negative  $g_{4}^{(1)}<0$ induces the phase 
$g_{2}^{(1)},g_{2}^{(2)},g_{4}^{(1)},g_{4}^{(2)} \to - \infty $.
 All these strong coupling fixed points correspond to the superconducting 
 phase of a fermionic two leg ladder, associated with different symmetries.\cite{varma_2chain,penc_2chain,fabrizio_2ch_rg,khveshenko_2chain,nagaosa_2ch,schulz_2chains,shelton_tj_ladder,balents_2ch,schulz_moriond}

\subsubsection{Three bands instabilities}
\label{sec:3bands-instability}
 \begin{figure*}[htbp]
\centerline{
\includegraphics[width=5.5cm]{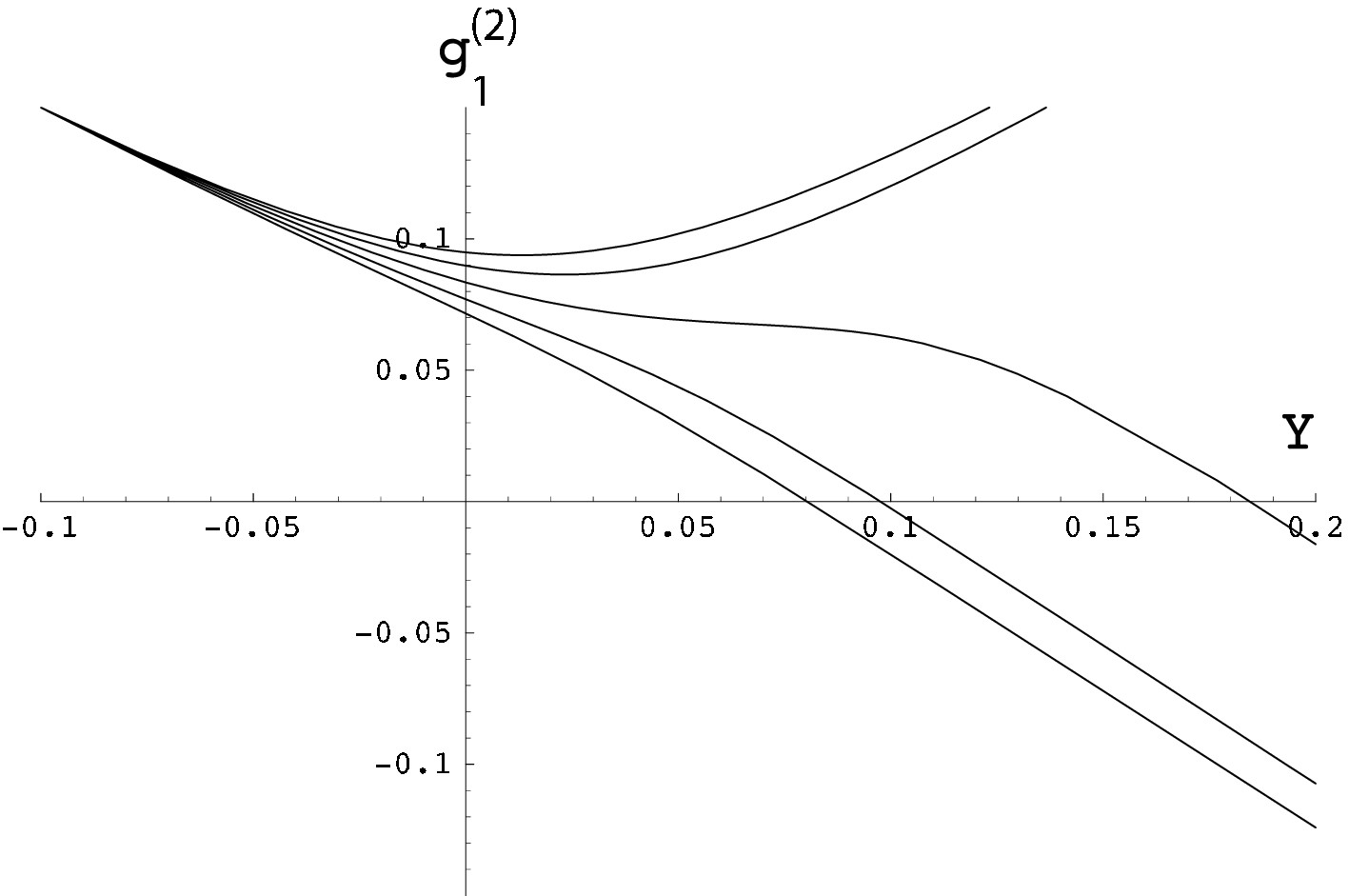}
\includegraphics[width=5.5cm]{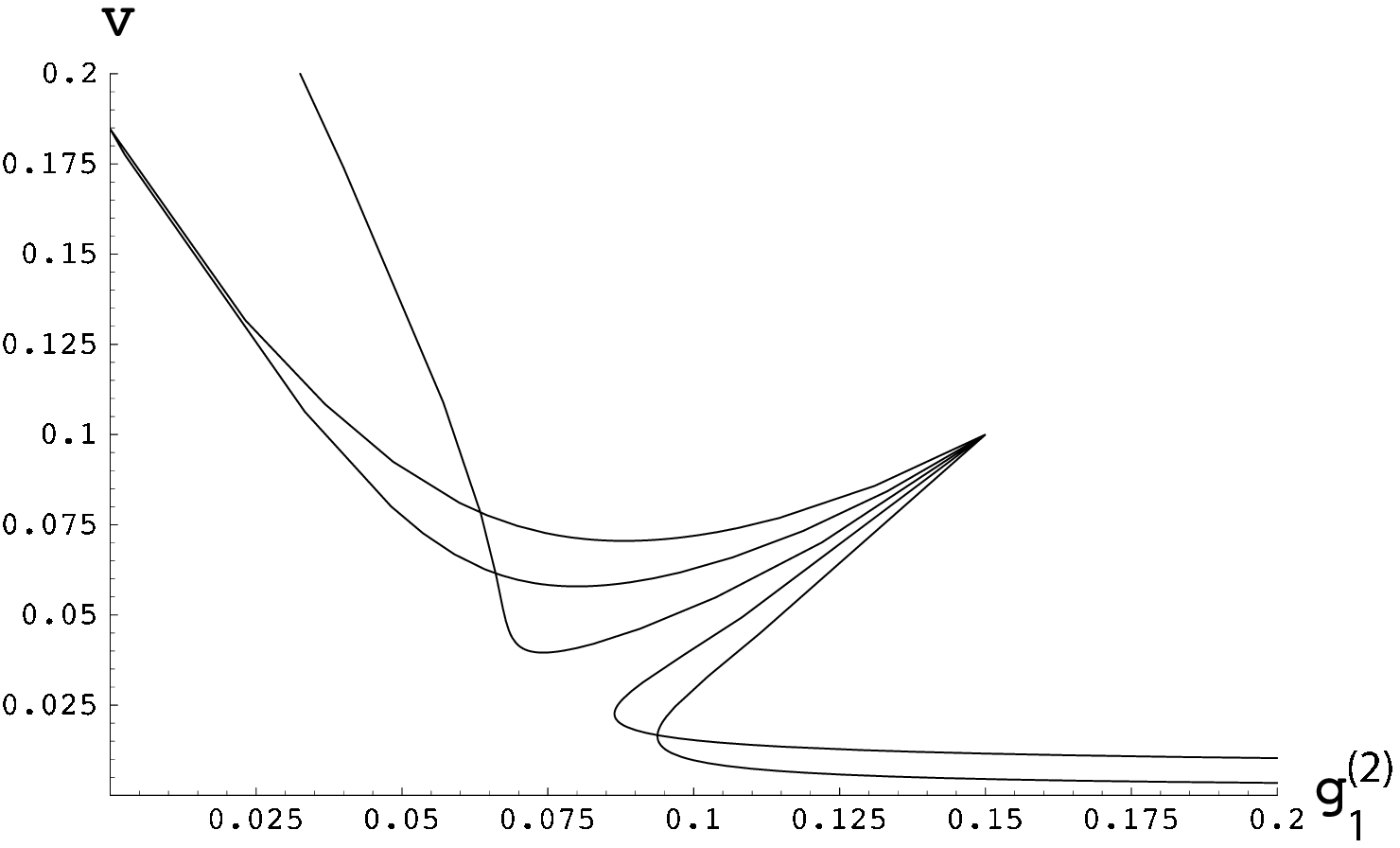}
\includegraphics[width=5.5cm]{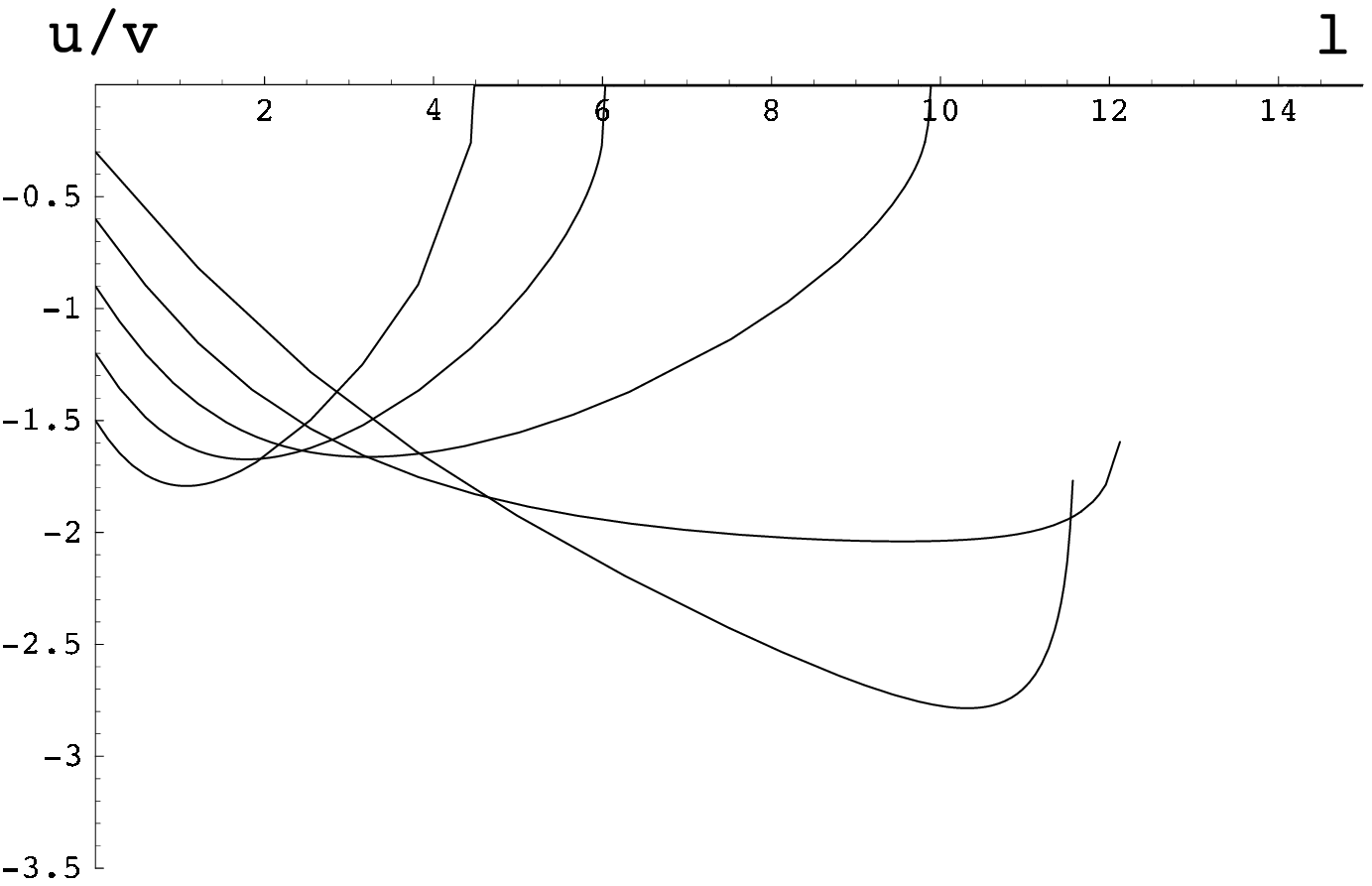}
}
\caption{Numerical renormalization group flow showing the instability
induced by a negative $u$ in the  
$Y=\tilde{g}_{4}^{(2)} - \tilde{g}_{2}^{(2)}, \tilde{g}_{1}^{(2)}$ plane (left), 
$\tilde{g}_{1}^{(2)},\tilde{v}$ plane (middle), and 
the ratio $u(l)/v(l)$ as a function of $l$ (right). These flow
corresponds to the  following initial values  
$\tilde{g}^{(1)} = 0.17,
\tilde{g}^{(2)} = 0.3,
\tilde{f}^{(1)}= 0.1,
\tilde{f}^{(1)}=  0.09,
\tilde{g}^{(1)}_{1} = 0.13,
\tilde{g}^{(2)}_{1} =  0.15,
\tilde{g}^{(1)}_{2} = 0.05,
\tilde{g}^{(2)}_{2} =   0.2,
\tilde{g}^{(1)}_{4} = 0.08,
\tilde{g}^{(2)}_{4} =   0.1$, 
and 
$\tilde{v} = 0.1$. $\tilde{u}$ was taken to negative values in step of
$0.03$ :   
$\tilde{u}= - 0.03 i$ for $i=1,5$.}
\label{fig:RG-u}
\end{figure*}

\begin{figure*}[htbp]
\centerline{
\includegraphics[width=5.5cm]{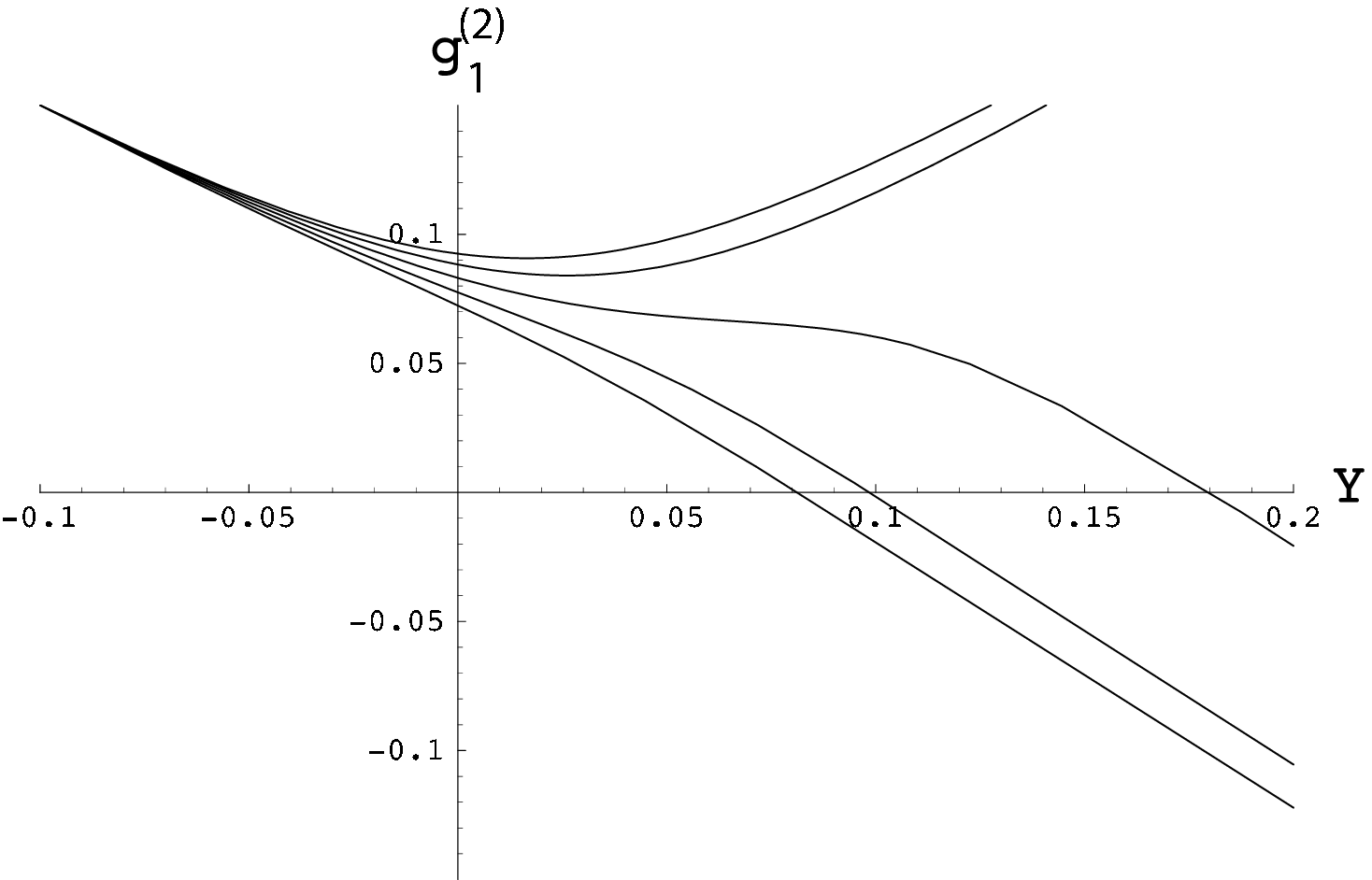}
\includegraphics[width=5.5cm]{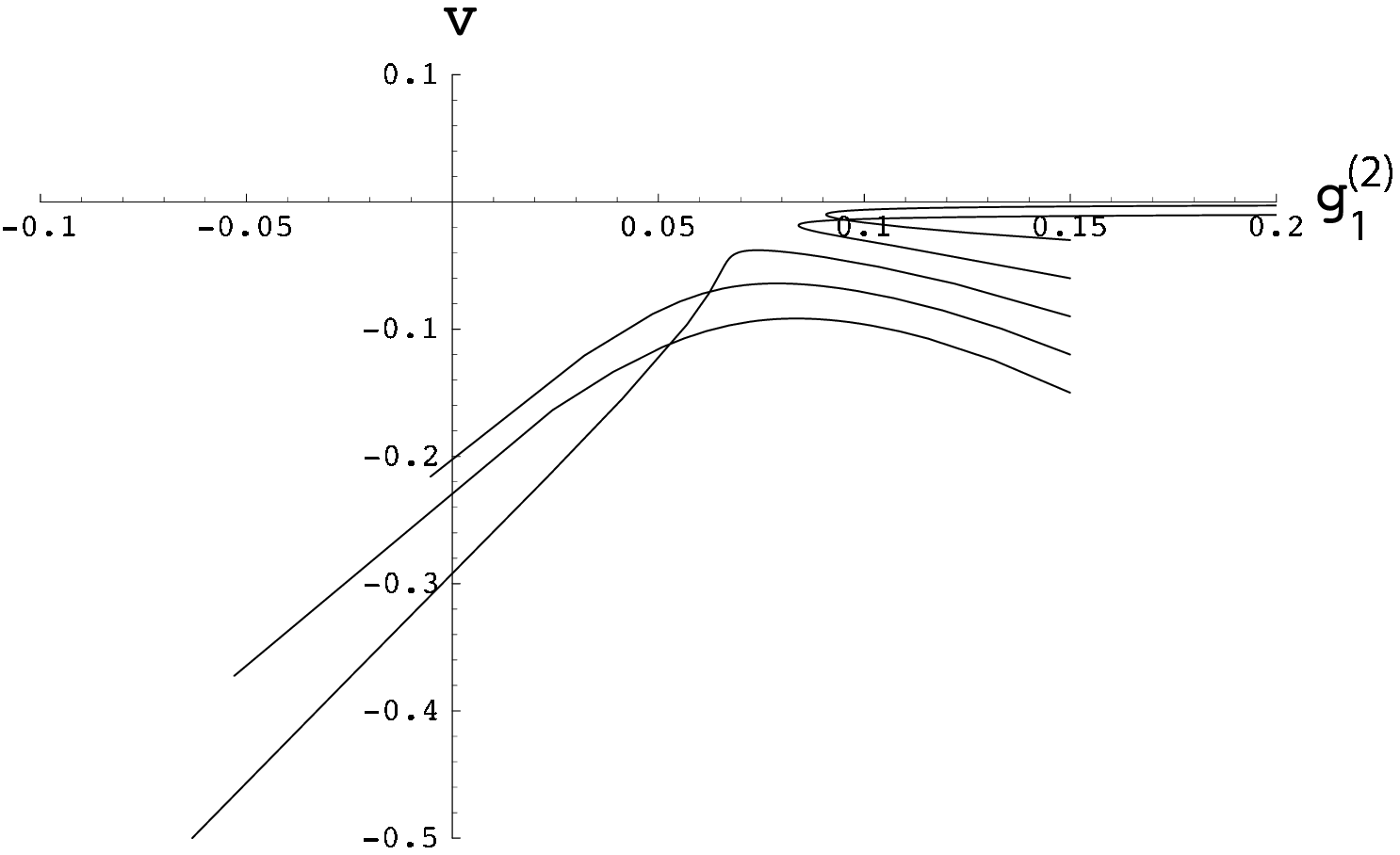}
\includegraphics[width=5.5cm]{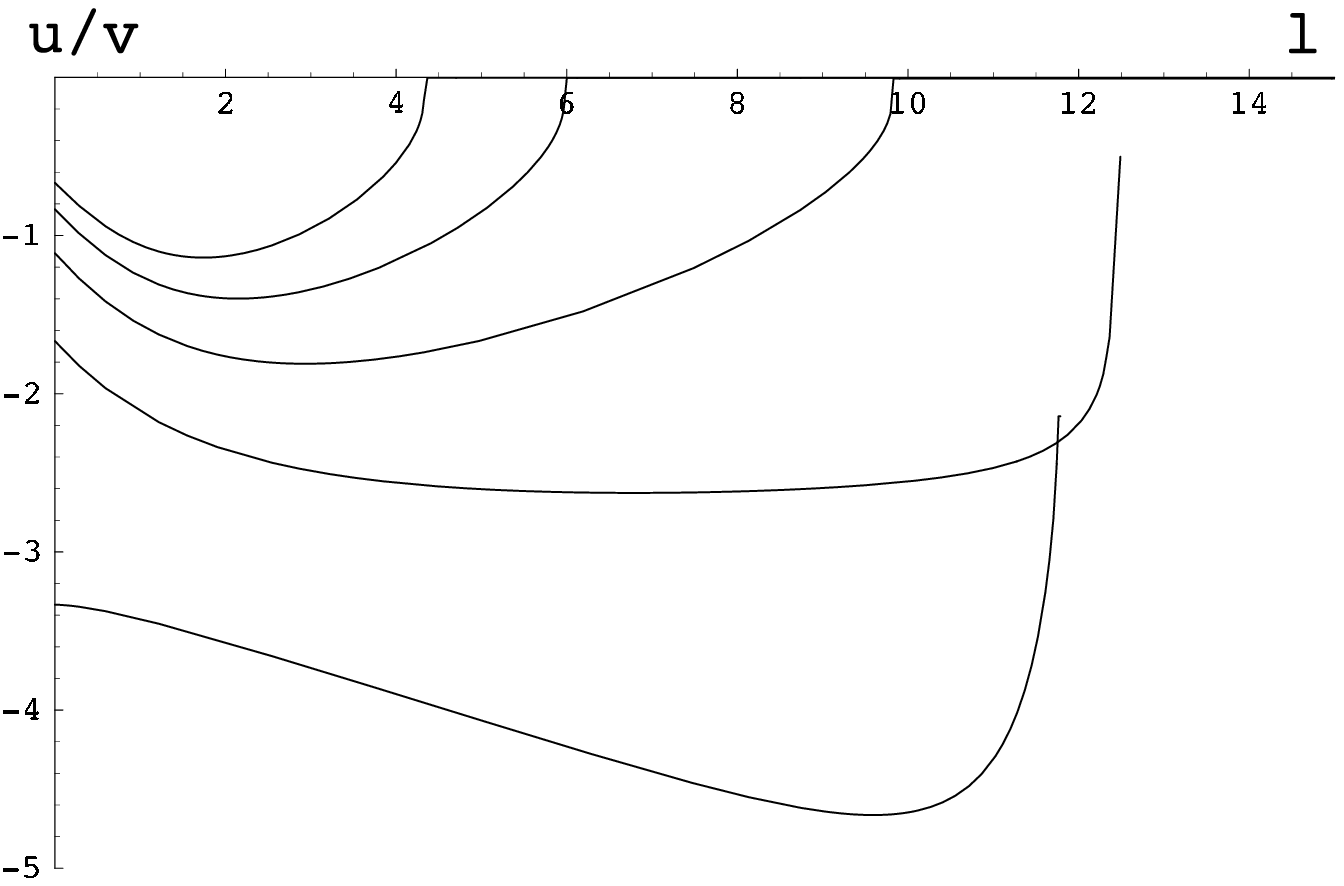}
}
\caption{Numerical renormalization group flow showing the instability
induced by a negative $v$ in the  
$Y=\tilde{g}_{4}^{(2)} - \tilde{g}_{2}^{(2)}, \tilde{g}_{1}^{(2)}$
plane (left), $\tilde{g}_{1}^{(2)},\tilde{v}$ plane (middle), and  
the ratio $u(l)/v(l)$ as a function of $l$ (right). These flow
corresponds to the  same initial values as in Fig.~\ref{fig:RG-u},
except that $\tilde{u}$ was held to $\tilde{u} = 0.1$ and $\tilde{v}=
- 0.03 i$ for $i=1,5$. }
\label{fig:RG-v}
\end{figure*}

 We now focus on the phases induced by attractive interactions
specific to the 3 bands nanotube model, namely $f^{(1)},u$ and $v$. In
all three cases, the dominant divergent couplings are 
$g_{2},f_{1},f_{2},g_{1}^{(2)},g_{2}^{(2)},g_{4}^{(2)},v$. We have
identified two pairs of asymptotic directions, one induced by negative
$f^{(1)}$ or 
$u$, and the second by $v$. These two pairs of strong coupling directions
differ only by the sign of the asymptotic $v (l^{*})$. In both cases,  
$g_{2},f_{1},f_{2},g_{2}^{(2)}$ flow towards $-\infty$,
$g_{4}^{(2)}$ towards $+\infty$ and $g_{1}^{(2)}\to \pm \infty$. 
The first direction corresponds to
$v\to \infty$, and the second to $v\to - \infty$. 
Two numerical flow obtained by slowly decreasing either $u$  
or $v$ to negative values are shown in figures~\ref{fig:RG-u} and
\ref{fig:RG-v}.   

\begin{figure}[htbp]
\centerline{\includegraphics[width=5cm]{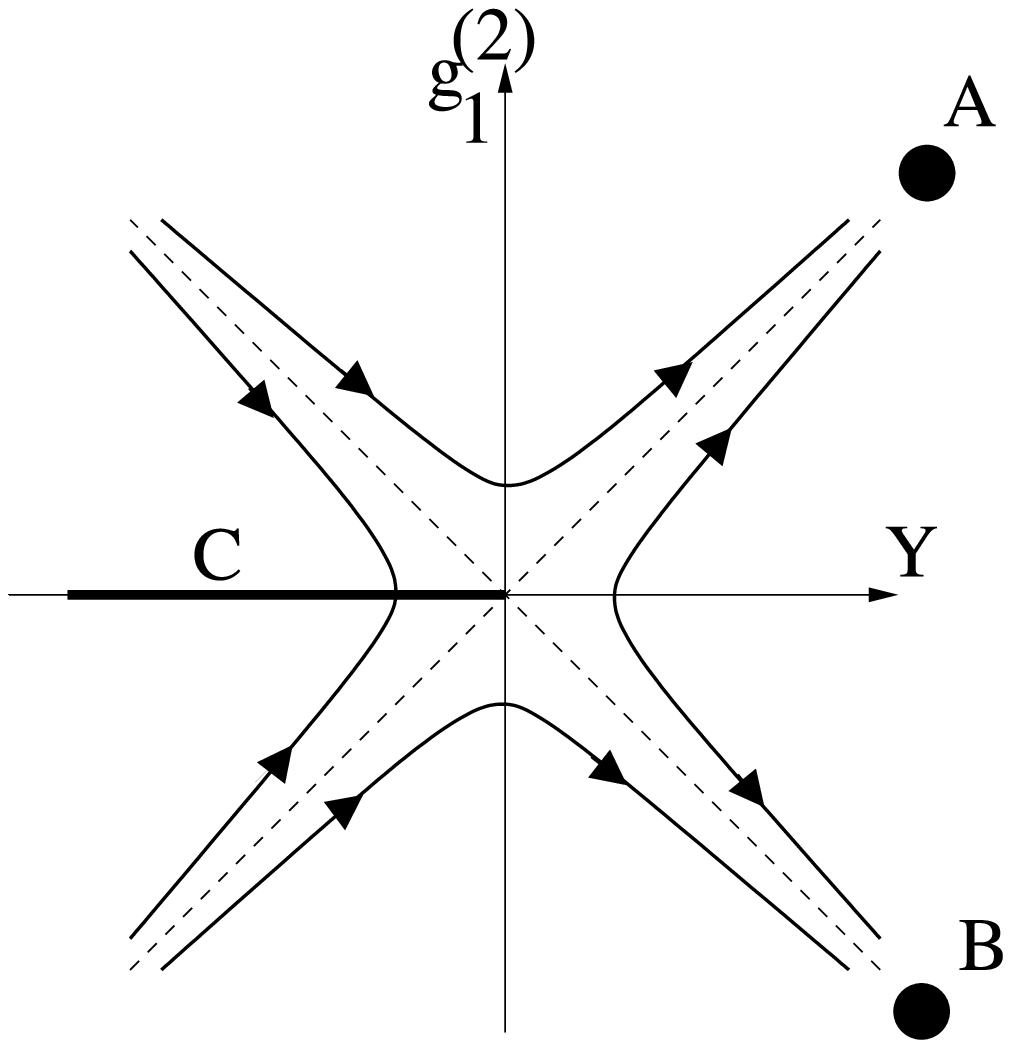}}
\caption{Schematic renormalization group flow in the 
$\tilde{g}_{1}^{(2)},Y=\tilde{g}_{4}^{(2)} - \tilde{g}_{2}^{(2)}$
plane, for $\tilde{v}=0$.} 
\label{fig:KT}
\end{figure}

 It is instructive to analyze further the renormalization flow by
focusing in the subspace of the dominant couplings 
$g_{2},f_{1},f_{2},g_{1}^{(2)},g_{2}^{(2)},g_{4}^{(2)},v$. Indeed
this subspace is stable under the RG equations (\ref{eq:RGeqs}). When
restricted to this subspace, these equations read 
\begin{subequations}
\label{eq:RGeqs-subspace}
\begin{align}
\partial_{l} \tilde{g}^{(2)} = &    - 2  \tilde{v}^2 \\
\partial_{l} \tilde{g}_{1}^{(2)} = &
    - 2 \tilde{g}_{1}^{(2)} \tilde{g}_{2}^{(2)} 
    + 2 \tilde{g}_{1}^{(2)} \tilde{g}_{4}^{(2)}
    - \tilde{v}^2  
\label{eq:scaling-g12}
\\
\partial_{l} \tilde{g}_{2}^{(2)} = & 
    -  (\tilde{g}_{1}^{(2)})^2
    -  \tilde{v}^2   \\
\partial_{l} \tilde{g}_{4}^{(2)} = &  (\tilde{g}_{1}^{(2)})^2   \\
\partial_{l} \tilde{f}^{(1)} = &     -2 (\tilde{f}^{(1)})^2 -2 \tilde{v}^2   \\
\partial_{l} \tilde{f}^{(2)} = &    -  (\tilde{f}^{(1)})^2    \\ 
\partial_{l} \tilde{v} =&
     - \left( 4 \tilde{f}^{(1)} 
     - 2 \tilde{f}^{(2)} 
     +  \tilde{g}_{1}^{(2)} 
     +  \tilde{g}^{(2)} 
     +  \tilde{g}_{2}^{(2)}\right) \tilde{v}
\end{align}
\end{subequations}
These equations possess two scaling invariants : 
$C = 2 \tilde{g}_{4}^{(2)}+2\tilde{g}_{2}^{(2)}-\tilde{g}^{(2)}$ and
$D = 2\tilde{f}^{(2)}-\tilde{f}^{(1)}+\tilde{g}^{(2)}$. 
Let us start by considering the RG flow in the subspace
$\tilde{v}=0$. Introducing  the variable $Y=\tilde{g}_{4}^{(2)} -
\tilde{g}_{2}^{(2)}$, the RG equations reduce to those of 
Kosterlitz and Thouless : 
 \begin{subequations}
\label{eq:RGeqs-subspace-2}
\begin{align}
\partial_{l} Y & = 2  (\tilde{g}_{1}^{(2)})^2 \quad ; \quad 
\partial_{l} \tilde{g}_{1}^{(2)}   =   2 \tilde{g}_{1}^{(2)}  Y \\
\partial_{l} \tilde{f}^{(1)} = &     -2 (\tilde{f}^{(1)})^2 
\end{align}
\end{subequations}
and $\tilde{g}^{(2)} $ is a flow constant. The asymptotic solutions
are thus the two directions (A and B on Fig.~\ref{fig:KT})
$Y(l)\simeq 1/(2(l^*-l))$ and $\tilde{g}_{1}^{(2)} = \pm
1/(2(l^*-l))$ and the line (C) $\tilde{g}_{1}^{(2)} = 0$, coupled to
the solutions $\tilde{f}^{(1)} =0$ or $\tilde{f}^{(1)} =-
1/(2(l^*-l))$ .
 The solutions corresponding to $\tilde{g}_{1}^{(2)} = 0$ or
$\tilde{f}^{(1)} =0$ are easily found to be unstable when
introducing a small $v$.

 The scaling behavior of $\tilde{v}$ can be deduced from inspection of
 equations (\ref{eq:RGeqs-subspace}).  Except for extremely large
 initial value of $\tilde{g}_{4}^{(2)}$, $\tilde{v}$ will always end
 up diverging to $\pm \infty$, in the direction given by its initial
 sign.
 The scaling equation (\ref{eq:scaling-g12}) shows that an increasing
 $v (l)$ leads to an instability of the asymptotic direction $A$. The only
 remaining scaling direction driven by $v$ corresponds to the point $B$ in
 Fig.~\ref{fig:KT}. The other possibility corresponds to strong
 initial  intra-band repulsion $g_{4}^{(2)}$, leading to a large and
 positive $g_1^{(2)}$. These strong
 interaction naturally forbid superconducting interactions, and we 
 will not consider them in the following. 

To complement the above
 analysis perturbative in $v$, we have analytically solved the
 equations
(\ref{eq:RGeqs-subspace}) with the ansatz $g (l) \simeq A_{g}/(l-l^*)$
for all couplings.  We have found that besides the $\tilde{v}=0$ fixed
points, there exist three strong couplings directions corresponding to
$\tilde{g}^{(2)}, \tilde{f}^{(1)}, \tilde{f}^{(2)},g_{2}^{(2)} \to
-\infty$,
$g_{4}^{(2)}\to +\infty $
 and
$\tilde{g}_{1}^{(2)}\to \pm \infty$ (together with $\tilde{v}\to
\pm\infty$).  The first direction corresponds to the direction
$\tilde{g}_{1}^{(2)} = 1/(2(l^*-l))$ (point A in
Fig.~\ref{fig:KT}) which is the point induced by a very strong
initial $g_{4}^{(2)}$ discussed above.  
Both the last two directions correspond to the same
limiting sign of the coupling constants. They differ only by the
numerical value of the ansatz parameters, and correspond to the
main instability discussed in the following.  
 We have thus identified a new instability, specific to the
3 bands model we consider. We now turn to the bosonization formalism
to identify the nature of the phase corresponding to this
  renormalization flow direction.

\section{Bosonization and Nature of the instabilities}
\label{sec:phases}

 The purpose of this section is to identify the nature of the
instability for  the  previously identified strong
coupling direction listed in table~\ref{table:subspace}. 
This will be achieved using the bosonization formalism within the
subspace corresponding to the dominant couplings. 
We will pay special attention to the proper definition of so-called 
Klein factor. 
We first start by defining our conventions on the non-interacting
three band model defined in section \ref{sec:model}.

\begin{table}
\begin{tabular}{|c|c|c|c|c|c|c| }
\hline
\multicolumn{5}{|c|}{back scattering} & \multicolumn{2}{c|}{forward scatt.} \\
\hline
$g^{(2)}$  & $f^{(2)}$ & $g_{1}^{(2)}$ & $g_{2}^{(2)}$ &
$g_{4}^{(2)}$ &  $f^{(1)}$ & $v$
\\
\hline  
$-\infty$ & $-\infty$ &$-\infty$ &$- \infty$ &$-\infty$ &$+\infty$
&$\pm \infty$ 
\\  
\hline
\end{tabular}
\caption{Table of the dominant couplings and their asymptotic directions
corresponding to the 3 bands instability on which we focus.} 
\label{table:subspace}
\end{table}

\subsection{The ``Condensed Matter'' Bosonization Dictionary}
\label{sec:CondensedMatter}

In the standard ``condensed matter'' bosonization procedure, we
represent the annihilation operators 
of right and left moving fermions, defined in (\ref{eq:def-psi}),   
as\cite{haldane_bosonisation,heidenreich_bosonisation,delft_bosonization}: 
\begin{subequations}\label{eq:bosonized-fermions}
\begin{align}
\psi_{R,m,\sigma}(x) &= \eta_{R,m,\sigma} \frac{1}{\sqrt{2\pi a}} 
 e^{-i \Phi_{R,m,\sigma}(x)}  \\
\psi_{L,m,\sigma}(x) &= \eta_{L,m,\sigma} \frac{1}{\sqrt{2\pi a}} 
e^{i \Phi_{L,m,\sigma}(x)}, 
\end{align}
\end{subequations}
\noindent where we introduced Majorana fermion operators (the
so-called Klein factors) $ \eta_{R/L,m,\sigma}$  
 that satisfy:
\begin{subequations}\label{eq:KleinConvention}
\begin{align}
\label{eq:majorana-anticom}
\{\eta_{R,m,\sigma},\eta_{R,m',\sigma'}\}_+ &= 
2\delta_{m,m'}\delta_{\sigma,\sigma'},
\\  
\{\eta_{L,m,\sigma},\eta_{L,m',\sigma'}\}_+ &= 
2\delta_{m,m'}\delta_{\sigma,\sigma'},
\\
\{\eta_{L,m,\sigma},\eta_{L,m',\sigma'}\}_+ &= 0.
\end{align}
\end{subequations}
 Note that in this convention, we introduce one Klein factor per
set of quantum numbers $(\pm k_{F},m,\sigma)$. 
With these anticommutation relations, the proper anticommutation relations
for the fermion operators defined in (\ref{eq:bosonized-fermions})
 are satisfied with  the following commutation
  relation of the fields $\Phi_{R/L}$ :
\begin{align}
\nonumber 
\lbrack\Phi_{R,m,\sigma}(x),\Phi_{R,m',\sigma'}(x')\rbrack &=
i\pi\delta_{m,m'}\delta_{\sigma,\sigma'}\mathrm{sign}(x-x') \\ 
\nonumber 
 \lbrack\Phi_{L,m,\sigma}(x),\Phi_{L,m',\sigma'}(x')\rbrack &=
-i\pi\delta_{m,m'}\delta_{\sigma,\sigma'}\mathrm{sign}(x-x') \\  
\label{eq:chiral-commutator}
 \lbrack\Phi_{L,m,\sigma}(x),\Phi_{R,m',\sigma'}(x')\rbrack &= 0. 
\end{align}
With these conventions, the bosonized
non-interacting Hamiltonian 
reads\cite{solyom_revue_1d,emery_revue_1d,haldane_bosonisation,%
           heidenreich_bosonisation,delft_bosonization}:   
\begin{equation}
H= \sum_{m=0,\pm 1 \atop \sigma=\uparrow,\downarrow}
\int \frac{dx}{4\pi} v_F \left[ (\nabla \Phi_{R,m,\sigma})^2+(\nabla
\Phi_{L,m,\sigma})^2\right],    
\end{equation}
Finally, the densities of right moving and left moving fermions read
respectively\cite{haldane_bosonisation,heidenreich_bosonisation,delft_bosonization}:
\begin{subequations}\label{eq:density-bosonized}
\begin{align}
 \rho_{R,m,\sigma}=-\frac{\nabla \Phi_{R,m,\sigma}}{2\pi} \\ 
  \rho_{R,m,\sigma}=-\frac{\nabla \Phi_{L,m,\sigma}}{2\pi} 
\end{align}
\end{subequations}


Keeping only the $g_2$ processes defined in section
\ref{sec:interactions},  we express the forward scattering part of
the interactions in terms of the above densities
(\ref{eq:density-bosonized}) : 
\begin{align} \nonumber 
 H_{\text{forward}} =
  &  g_4^{(2)} \sum_{\sigma,\sigma'} (\rho_{R,1,\sigma}
\rho_{L,1,\sigma'}+\rho_{R,-1,\sigma} \rho_{L,-1,\sigma'})  \\
\nonumber  
+& f^{(2)} \sum_{\sigma,\sigma'}
[(\rho_{R,1,\sigma}+\rho_{R,-1,\sigma}) \rho_{L,0,\sigma'} \\
\nonumber  
& \quad \quad \quad + (\rho_{L,1,\sigma}+\rho_{L,-1,\sigma})
\rho_{R,0,\sigma'} ] \nonumber \\ \nonumber  
 + & g_2^{(2)} \sum_{\sigma,\sigma'} 
(\rho_{R,1,\sigma} \rho_{L,-1,\sigma'}+  \rho_{R,-1,\sigma} \rho_{L,1,\sigma'}) \\  
 + & g^{(2)} \sum_{\sigma,\sigma'} \rho_{R,0,\sigma} \rho_{L,0,\sigma'} 
\end{align}
 Using (\ref{eq:density-bosonized}), these expressions can be reduced
to quadratic expressions in  the fields $\Phi_{R/L,m,\sigma}$. 
 Note that we will treat the $g_{1}^{(2)}$ term below with the
 backscattering part of the Hamiltonian : although it appears as a
 forward scattering term, it cannot be reduced to a density-density
 coupling and its treatment closely follows the one for the
 backscattering couplings.

\subsection{Backscattering Interactions} 

\subsubsection{The Klein factors problem}

 Whereas the bosonized forward scattering part of the Hamiltonian is 
 function solely of the densities, and thus 
does not depend on the convention chosen for the Klein factor
(\ref{eq:KleinConvention})  and fields (\ref{eq:chiral-commutator}),
the situation is different for the back-scattering part of the
action (\ref{eq:formal-Sint}). Quite generally, these back-scattering
operators can be written as 
\begin{multline}\label{eq:formalBackScatter}
 \psi^{\dagger}_{R,m_{1},\sigma} \psi^{\dagger}_{L,m_{2},\sigma'}
\psi_{L,m_{3},\sigma'}\psi_{R,m_{4},\sigma}\\
= \frac{1}{(2\pi a)^{2}}
\mathcal{N}_{\{m \},\sigma,\sigma'}
\mathcal{O}_{\{m \},\sigma,\sigma'}
\end{multline}
where we assumed the transverse momentum conservation
$m_{1}+m_{2}=m_{3}+m_{4}$ and we defined the product of Majorana
fermions 
\begin{equation}
\mathcal{N}_{\{m \},\sigma,\sigma'} =  
\eta_{R,m_{1},\sigma} \eta_{L,m_{2},\sigma'}
\eta_{L,m_{3},\sigma'}\eta_{R,m_{4},\sigma}
\end{equation}
 and the product of vertex operators 
\begin{equation}
\mathcal{O}_{\{m \},\sigma,\sigma'} = 
e^{i \Phi_{R,m_{1},\sigma}}
e^{-i \Phi_{L,m_{2},\sigma'}}
e^{i \Phi_{L,m_{3},\sigma'}}
e^{- i \Phi_{R,m_{4},\sigma}}. 
\end{equation}
The usual strategy is  to find a representation such that the
operators $\mathcal{O}_{\{m \},\sigma,\sigma'} $ commute with each
other, and similarly for the products of 
four Majorana fermions $\mathcal{N}_{\{m \},\sigma,\sigma'}$. 
It is
important to note  that the only case discussed in
Ref.~\onlinecite{schulz_moriond} is the one in which all the vertex
operators $\mathcal{O}_{\{m \},\sigma,\sigma'} $ 
are already commuting so that no redefinition of  the fields is necessary. 
  When determining the ground state, it is then
possible to consider the Sine-Gordon form of the Hamiltonian,
obtained by replacing the operators 
$\mathcal{N}_{\{m\},\sigma,\sigma'}$ by their eigenvalues. 

 Obviously, while the $4$ fermions operators
(\ref{eq:formalBackScatter}) always commute with each other,  the
above condition of independent commutation of the 
$\mathcal{N}_{\{m \},\sigma,\sigma'}$ and 
$\mathcal{O}_{\{m \},\sigma,\sigma'} $ becomes more and more difficult
to fulfill with an increasing number of fermion species. This is
particularly true for our $3$ bands model, corresponding to $12$ fermionic
species ($2$ spins and $6$ Fermi points), and we can  check on the
bosonized expressions derived in App.~\ref{app:bosonization},
Eqs. (\ref{eq:g12-condmat}), (\ref{eq:f1-condmat}) and
(\ref{eq:v-condmat}),  that
this condition cannot be satisfied for the operators we consider
within the convention defined by Eqs.~(\ref{eq:bosonized-fermions},\ref{eq:KleinConvention},%
\ref{eq:chiral-commutator}).  
 Indeed, the products of four Majorana fermion operators 
are commuting when they have an even number of Majorana fermions in common
and anticommuting otherwise.
In the second case, which occurs for our model (see {\it e.g} the
operator $v$), the corresponding operators 
$\mathcal{O}_{\{m \},\sigma,\sigma'}$ also contains an odd number 
 of vertex operators in common.  Since the vertex 
operators associated with fermions are anticommuting when they correspond
to the same fermion species, we recover in the bosonization formalism
the commutation of the four Fermi operators, but the independent
commutations of the $\mathcal{O}_{\{m \},\sigma,\sigma'} $ and
$\mathcal{N}_{\{m \},\sigma,\sigma'}$ is not possible. 
We thus need to change our bosonization
convention for this particular model.     

\subsubsection{``Field Theory'' Convention}
\label{sec:FieldTheory}

The problem we have to deal with is thus whether it is possible to redefine
the Majorana fermion operators (\ref{eq:KleinConvention}) and the
commutation relations of the fields (\ref{eq:chiral-commutator}) 
that appear in the vertex operators in such a way  
that all the new products of four Majorana fermion
operators are commuting with each other  and simultaneously all the new 
products of vertex operators are also commuting with each other.
Another possible convention, different 
from the above ``condensed matter'' convention (left/right moving 
chiral field of the same band commuting with each other, and 
one Majorana fermion per Fermi point\cite{delft_bosonization}),
 consists in what we will call   
the ``quantum field theory'' convention. We now introduce a 
single Majorana fermion for a pair of right and left fermion which
need not have the same quantum numbers (same specie). Correspondingly,
the chiral fields for this pair 
have a nonzero commutator. 
 In this representation, the fermion operators are now expressed
 as\footnote{Note that we could slightly extend this convention to the case of spin-dependent interactions by allowing the Left and Right fermions of the pair to have different spins.}  
\begin{subequations}\label{eq:bosonized-fermions2}
\begin{align}
\psi_{R,m,\sigma} (x) &= \frac 1 {\sqrt{2\pi\alpha}} 
   e^{-i\tilde{\Phi}_{R,m,\sigma}(x)}  \eta_{m,\sigma}, \\
\psi_{L,P (m),\sigma} (x) &= \frac 1 {\sqrt{2\pi\alpha}} 
   e^{i\tilde{\Phi}_{L,m,\sigma}(x)} \eta_{m,\sigma},
\end{align}
\end{subequations}
 where $P$ is a permutation of the band indices (fermion species). 
Now the field commutations relations are modified into :  
\begin{align}
\lbrack \tilde{\Phi}_{R,m,\sigma}(x),
        \tilde{\Phi}_{R,m',\sigma'}(x')\rbrack &=
i\pi \delta_{m,m'} \delta_{\sigma,\sigma'} \mathrm{sign}(x-x'),  
\nonumber \\
\lbrack\tilde{\Phi}_{L,m,\sigma}(x),\tilde{\Phi}_{L,m',\sigma'}(x')\rbrack
& =
-i\pi \delta_{m,m'} \delta_{\sigma,\sigma'} \mathrm{sign}(x-x'),
\nonumber \\
\label{eq:chiral-commutator2}
\lbrack\tilde{\Phi}_{R,m,\sigma}(x),\tilde{\Phi}_{L,m',\sigma'}(x')\rbrack
& =
i\pi \delta_{m,m'} \delta_{\sigma,\sigma'}
\end{align}
and the Majorana fermion operators $\eta_{m,\sigma}$ satisfy:
\begin{equation}\label{eq:KleinConvention-qft}
  \{\eta_{m,\sigma},\eta_{m',\sigma'}\}=2\delta_{m,m'}\delta_{\sigma,\sigma'} .
\end{equation}
We discuss the equivalence of these two representations in the
appendix~\ref{app:klein}. Let us now apply this convention to the
present model. We have found that the suitable (necessary)
permutation $P$ of band indices in Eq.~(\ref{eq:bosonized-fermions2})
simply permutes the bands $+1$ and $-1$. 

The remaining interactions $g_{1}^{(2)},f^{(1)},v$ are conveniently
expressed in terms of the 
following non chiral fields~\cite{schulz_houches_revue}: 
\begin{eqnarray}
\label{eq:theta-phi-def}
\theta_{m,\sigma} &=& \frac 1 2 
(\tilde{\Phi}_{L,m,\sigma} - \tilde{\Phi}_{R,m,\sigma}),
\nonumber  \\   
\phi_{m,\sigma}   &=& \frac 1 2
(\tilde{\Phi}_{L,m,\sigma} + \tilde{\Phi}_{R,m,\sigma}).  
\end{eqnarray}
These fields satisfy 
$[\phi_{m,\sigma}(x),\phi_{m',\sigma'}(x')]=
[\theta_{m,\sigma}(x),\theta_{m',\sigma'}(x')]=0$ and
$[\phi_{m,\sigma}(x),\theta_{m',\sigma'}(x')]=
i \frac{\pi} 2 \delta_{m,m'}\delta_{\sigma,\sigma'}
\textrm{sign}(x'-x)$. 
Taking the
derivative with respect to $x'$ and introducing $\Pi_{m,\sigma}(x)=\frac 1 \pi
\partial_x \theta_{m,\sigma}$ one finds
$[\phi_{m,\sigma}(x),\Pi_{m,\sigma}(x')]=i\delta(x-x')$, showing  
that the fields $\Pi_{m,\sigma}$ and $\phi_{m,\sigma}$ are canonically conjugate. 
It is convenient to introduce the total and relative ``charge'' and ``spin''
fields\cite{schulz_houches_revue}:
\begin{subequations}
\begin{align}
  \theta_{c,m}=\frac 1 {\sqrt{2}} (\theta_{m,\uparrow} +\theta_{m,\downarrow}), \\ 
 \theta_{s,m}=\frac 1 {\sqrt{2}} (\theta_{m,\uparrow} -\theta_{m,\downarrow}), \\ 
 \phi_{c,m}=\frac 1 {\sqrt{2}} (\phi_{m,\uparrow} +\phi_{m,\downarrow}), \\ 
 \phi_{s,m}=\frac 1 {\sqrt{2}} (\phi_{m,\uparrow} -\phi_{m,\downarrow}). 
\end{align}
\end{subequations}
Finally, the following rotation for the charge modes will simplify
the expressions of the interactions : 
\begin{equation}\label{eq:charge-rot}
  \left(\begin{array}{c} \phi_{c,+} \\ \phi_{c,A} \\
  \phi_{c,B} \end{array} \right) = \left(\begin{array}{ccc} \frac 1
  {\sqrt{3}} & \frac 1 {\sqrt{3}}& \frac 1 {\sqrt{3}} \\ \frac 1
  {\sqrt{2}} & -\frac 1 {\sqrt{2}} & 0 \\ \frac 1 {\sqrt{6}} & \frac 1
  {\sqrt{6}}& -\frac 2 {\sqrt{6}} \\ \end{array}\right)
  \left( \begin{array}{c} \phi_{c,1} \\ \phi_{c,-1} \\
  \phi_{c,0} \end{array}\right). 
\end{equation}
The same rotation is performed for $\theta_{c,i}$ so that
 canonical commutation relations are preserved. A different rotation
 must be performed for the
spin modes:
\begin{equation}
  \label{eq:spin-rot} \left(\begin{array}{c} \phi_{s,+} \\
  \phi_{s,-} \end{array} \right) = \left(\begin{array}{cc} \frac 1
  {\sqrt{2}} & \frac 1 {\sqrt{2}} \\ \frac 1 {\sqrt{2}} & -\frac 1
  {\sqrt{2}} \end{array} \right) \left(\begin{array}{c} \phi_{s,1} \\
  \phi_{s,-1} \end{array} \right)
\end{equation}
 With these notations and within the above field-theoretic
 representation,  the bosonized expression of the $g_1^{(2)}$ term
 becomes:
\begin{multline}
\label{eq:g12-field-theory}
\frac{g_1^{(2)}}{(\pi a)^2} \int dx 
\cos 2 \theta_{c,A}  \biggl( 
\cos 2\theta_{s-} \\
+ \cos 2\phi_{s-}
\eta_{1,\uparrow}\eta_{1,\downarrow}\eta_{-1,\downarrow}\eta_{-1,\uparrow}
\biggr). 
\end{multline}
After some simple algebra, we find for the $f^{(1)}$ part of the
Hamiltonian the simplified expression 
\begin{widetext}
\begin{multline}
-\frac{ f^{(1)}} {(2\pi a)^2} \sum_\sigma \biggl\{
 e^{-i \sigma \sqrt{2} (\theta_{s,0} -\phi_{s,0})} 
\eta_{0,\sigma} \eta_{0,-\sigma} 
e^{i\sigma(\theta_{s,+}+\phi_{s,+})} 
\left[
e^{i\sigma(\theta_{s,-}+\phi_{s,-})} 
\eta_{-1,-\sigma} \eta_{-1,\sigma} 
+ e^{-i\sigma(\theta_{s,-}+\phi_{s,-})}
\eta_{1,-\sigma} \eta_{1,\sigma} \right]
\\
 +  e^{-i \sigma \sqrt{2} (\theta_{s,0} + \phi_{s,0})}
\eta_{0,\sigma} \eta_{0,-\sigma}
e^{i\sigma(\theta_{s,+}-\phi_{s,+})} 
\left[
e^{i\sigma(\theta_{s,-}-\phi_{s,-})} \eta_{1,-\sigma} \eta_{1,\sigma} 
+ e^{-i\sigma(\theta_{s,-}-\phi_{s,-})} 
\eta_{-1,-\sigma} \eta_{-1,\sigma}
\right]
\biggr\}, 
\end{multline}
 and for the only remaining $v$ coupling, the bosonization
 expressions reads 
\begin{multline}
\frac {2 v}{(\pi a)^2} \Biggl\{
\eta_{1\uparrow}\eta_{1\downarrow}
\eta_{0\downarrow}\eta_{0\uparrow}
\biggl[
\cos (\sqrt{3}\theta_{cB}) \cos \theta_{cA} 
\cos\phi_{s-} \cos (\sqrt{2}\phi_{s0}+\phi_{s+}) 
+ \sin (\sqrt{3}\theta_{cB}) \sin \theta_{cA} 
\sin\phi_{s-} \sin (\sqrt{2}\phi_{s0}+\phi_{s+}) 
\biggr] \\
- \biggl[
\cos (\sqrt{3}\theta_{cB}) \cos \theta_{cA} 
\cos\theta_{s-} \cos (\sqrt{2}\theta_{s0}-\theta_{s+}) 
- \sin (\sqrt{3}\theta_{cB}) \sin \theta_{cA} 
\sin\theta_{s-} \sin (\sqrt{2}\theta_{s0}-\theta_{s+})
\biggr]  \Biggl\}
\end{multline}
\end{widetext}
We immediately observe that the change of bosonization convention 
results in an  important simplification. 
First,  in this field-theoretic representation, 
the Majorana fermion product in the $g_1^{(2)}$
term is now commuting with the Majorana fermion products that appear in 
the $f^{(1)}$ and the $v$ terms. This allows for a simultaneous
diagonalization of all these remaining Majorana fermions products.
Representing the 2 Majorana fermions products as pseudo-spins :  
$\eta_{1,\uparrow}\eta_{1,\downarrow} = i\tilde{\sigma}_{1}^{z}$, 
$\eta_{-1,\uparrow}\eta_{-1,\downarrow} = i\tilde{\sigma}_{-1}^{z}$, 
$\eta_{0,\uparrow}\eta_{0,\downarrow} = i\tilde{\sigma}_{0}^{z}$,
 and choosing the $+1$ eigenvalues of the $\tilde{\sigma}_{m}^{z}$, 
we obtain the final bosonized action, which takes the Sine-Gordon like
form. The first $g_{1}^{(2)}$ term reads 
\begin{equation}\label{eq:g12-final}
\frac{g_1^{(2)}}{(\pi a)^2} \int dx 
\cos 2 \theta_{c,A}  \biggl( 
\cos 2\theta_{s-} + \cos 2\phi_{s-}
\biggr),  
\end{equation}
the $f^{(1)}$ term simplifies into 
\begin{widetext}
\begin{multline}\label{eq:f1-final}
-\frac{ f^{(1)}} {(\pi a)^2} \biggl\{
\cos \left[(\theta_{s,+}+\phi_{s,+}) -\sqrt{2} (\theta_{s,0} -\phi_{s,0})\right]
\cos (\theta_{s,-}+\phi_{s,-})  \\
+ \cos \left[(\theta_{s,+}-\phi_{s,+}) -\sqrt{2} (\theta_{s,0} +\phi_{s,0})\right]
\cos (\theta_{s,-}-\phi_{s,-})
 \biggl\}, 
\end{multline}
and finally the $v$ term can be written as 
\begin{multline}\label{eq:v-final}
\frac {2 v}{(\pi a)^2} \Biggl\{
\cos (\sqrt{3}\theta_{cB}) \cos \theta_{cA} 
\cos\phi_{s-} \cos (\sqrt{2}\phi_{s0}+\phi_{s+}) 
+ \sin (\sqrt{3}\theta_{cB}) \sin \theta_{cA} 
\sin\phi_{s-} \sin (\sqrt{2}\phi_{s0}+\phi_{s+}) 
\\
\cos (\sqrt{3}\theta_{cB}) \cos \theta_{cA} 
\cos\theta_{s-} \cos (\sqrt{2}\theta_{s0}-\theta_{s+}) 
- \sin (\sqrt{3}\theta_{cB}) \sin \theta_{cA} 
\sin\theta_{s-} \sin (\sqrt{2}\theta_{s0}-\theta_{s+})
\Biggl\}
\end{multline}
\end{widetext}

\subsection{Analysis of the strong coupling fixed points :
superconducting instability}

 Having obtained the above expressions  
Eqs.~(\ref{eq:g12-final},\ref{eq:f1-final},\ref{eq:v-final}), we are now
ready to characterize the phases corresponding to the
strong coupling directions identified in the renormalization study
of Sec.~\ref{sec:3bands-instability}. We will only focus on the
new instability specific to a three band model. 
 This instability corresponds to a divergence of 
$g_{1}^{(2)}\to -\infty$, $f^{(1)}\to +\infty$ and $v\to \pm \infty$ 
(see table \ref{table:subspace}). The two signs of $v$ are possible,
depending on the driving attractive perturbations 
({\it e.g} $u$ or $v$).
From (\ref{eq:g12-final}), we find  that large negative values of
$g_{1}^{(2)}$ induce a locking of the field 
$\theta_{c,A}=0$. $\theta_{s-}$ and $\phi_{s-}$ being dual to each
other,  no further information on the spin part can be gained at this
point. The $f^{(1)}$ being a pure spin-current interaction part, we
will postpone its analysis to the next section. And finally, plugging 
the result  $\theta_{c,A}=0$ into (\ref{eq:v-final}), we find that
large values of $v$ will induce a locking of the charge field 
$\theta_{c,B}$ to
\begin{align}\label{}
\theta_{c,B} = \left\{
\begin{array}{l}
\frac{\pi}{\sqrt{3}} \textrm{ if } v>0\\
0 \textrm{ if } v < 0
\end{array}
 \right.
\end{align}
 Thus we find that in the ground state corresponding to this
 instability, the fields  $\phi_{c,A}$ and $\theta_{c,B}$ are locked 
so as to minimize the condensation energy. This implies that the corresponding
charge degrees of freedom develop a gap. The total charge remains gapless
as a result of the global $U(1)$ symmetry. Thus, only a single charge
degree of freedom remains. The analysis of the spin degrees of freedom
is more difficult, and is done in the next section. However, 
knowing which charge modes are gapped already
enables us to determine some of the order parameters, and find the
corresponding nature of the instability.  

In any case, the long range ordering of the charge fields
$\theta_{c,B}$ and $\phi_{c,A}$ has important consequences.  
Indeed, it is seen from the bosonized expressions (\ref{eq:CDW1}-\ref{eq:CDW2})
of the charge density wave operators that none of them can develop quasi-long range order.
On the other hand, superconducting fluctuations are strongly
reinforced by the ordering of the charge fields, as can be seen on
the corresponding expressions (\ref{eq:superconductivity}) derived in
the appendix \ref{app:bosonization}. Hence we have analyzed the
new instability as being driven by  superconducting fluctuations. 

It is worthwhile to contrast our results with those previously obtained
on three-leg ladders.\cite{arrigoni_3chain} In our notations, it was
found that in the three leg ladder system, the only charge field
developing long range order was $\theta_{c,A}$. Here, by contrast, we
find that two charge fields are developing a long range order,
$\theta_{c,A}$ and $\theta_{c,B}$. The difference of behavior of the
three-band nanotube and the three-leg ladder is a consequence of the
equality of the Fermi wavevectors of the band of angular momentum $\pm
1$ which is itself a consequence of the rotational symmetry of the
tube. This equality of wavevector allows extra interactions between
the bands $\pm 1$ such as $g_1^{(2)}$ and between the two bands $\pm
1$ and the band $0$ (such as $u$ or $v$). The existence of these
interactions is driving the system to a new fixed point. As one more
charge modes is gapped in the nanotube compared with the three leg
ladder, the reinforcement of superconducting fluctuations is expected
to be a stronger effect in the nanotube.

\subsection{Effective low energy spin theory for the instabilities}
\label{sec:NonAbelian}

The charge modes of the nanotubes being gapped, apart from the global
decoupled charge mode,  the corresponding low energy description of
the instability we consider consists only of the spin modes.  
Further progress in the understanding of this theory 
can be made by introducing 
the pseudo-fermion creation and annihilation  operators: 
\begin{align}\label{eq:psefofermions}
\Psi_{R,+}=\frac{\eta_+}{\sqrt{2\pi\alpha}}  e^{i(\theta_{s,+}-\phi_{s,+})}, \nonumber
\\
\Psi_{L,+}=\frac{\eta_+}{\sqrt{2\pi\alpha}}  e^{i(\theta_{s,+}+\phi_{s,+})}, \nonumber
\\
\Psi_{R,-}=\frac{\eta_-}{\sqrt{2\pi\alpha}}  e^{i(\theta_{s,-}-\phi_{s,-})}, \nonumber
\\
\Psi_{L,-}=\frac{\eta_-}{\sqrt{2\pi\alpha}}  e^{i(\theta_{s,-}+\phi_{s,-})},
\end{align}
\noindent and the associated Majorana fermion operators ($\nu=R,L$): 
\begin{eqnarray}
 \Psi_{\nu,+}&=&\frac 1 {\sqrt{2}} (-\zeta_{\nu,1}-i \zeta_{\nu,2}), \nonumber
 \\ 
\Psi_{\nu,-}&=&\frac 1 {\sqrt{2}} (\zeta_{\nu,3}+i \zeta_{\nu,0}). 
\end{eqnarray}
The interaction term proportional to  $g_1^{(2)}$ is then rewritten as:
\begin{equation}\label{eq:majorana-mass-0} 
 2 i \frac{g_1^{(2)}}{\pi a} \int dx \cos 2 \theta_{c,A} \zeta_{R,0}
 \zeta_{L,0} 
\end{equation}
we see that the interaction term proportional 
to $g_1^{(2)}$  gives a non-zero mass to the
$\zeta_{R/L,0}$ Majorana fermions, 
while leaving the  $\zeta_{R/L,3}$  fermions massless. This Ising
criticality is a consequence of the self-dual\cite{jose_planar_2d,lecheminant02_sdsg} character of the
interaction~(\ref{eq:g12-final}). In the context of two-leg ladders,
these self-dual interactions have been discussed in
Refs.~\onlinecite{finkelstein_2ch,schulz_2chains}.   

Moreover, the interaction term proportional to $f^{(1)}$
can also be reexpressed in terms of the fermion fields
$\zeta_{R/L,1,2,3}$. Indeed, we have the relations:
\begin{eqnarray}
  \label{eq:majorana-currents}
  \frac{e^{-i(\theta_{s+}+\phi_{s+})}}{\pi \alpha} \cos (\theta_{s-}+\phi_{s-}) &=& -i
  (\zeta_{L,2} \zeta_{L,3}+i \zeta_{L,3} \zeta_{L,1}),  \nonumber \\ 
   \frac{e^{-i(\theta_{s+}-\phi_{s+})}}{\pi \alpha}  \cos (\theta_{s-}-\phi_{s-}) &=& -i
  (\zeta_{R,2} \zeta_{R,3}+i \zeta_{R,3} \zeta_{R,1}).\nonumber \\   
\end{eqnarray}
In fact, this representation is well
known\cite{zamolodchikov_fateev} and has been used to study the
two-leg spin ladder\cite{shelton_spin_ladders,allen} and the
two-channel Kondo effect\cite{emery_2channel,delft_bosonization}. 
Using the equivalence between Majorana fermions and the two
dimensional Ising model, it is also possible to reexpress the interaction $v$ 
using order and disorder parameters of the quantum Ising
model\cite{kadanoff_gaussian_model,zuber_77,schroer_ising,%
ogilvie_ising,boyanovsky_ising}. Indeed, one has the
relations\cite{nersesyan01_ising_review}: 
\begin{eqnarray}
  \label{eq:double-ising}
  \cos \phi_{s+}&=&\mu_1 \mu_2, \nonumber \\ 
  \cos \theta_{s+}&=&i \kappa_1 \sigma_1 \mu_2,  \nonumber \\ 
  \sin \theta_{s+}&=&-i \kappa_2 \mu_1 \sigma_2,  \nonumber \\
  \sin \phi_{s+}&=&-i \kappa_1 \kappa_2 \sigma_1  \sigma_2, 
\end{eqnarray}
\noindent and similar relations for $\phi_{s-}$ and $\theta_{s-}$ with
  $(\mu_1,\sigma_1) \to (\mu_3,\sigma_3)$ and $(\mu_2,\sigma_2) \to
  (\mu_0,\sigma_0)$. With these relations, we easily find that:
  \begin{eqnarray}
    \label{eq:ising-su2-2}
    \cos \phi_{s-} e^{i\phi_{s+}} &=&\mu_3 \mu_0 [\mu_1 \mu_2 + \kappa_1
    \kappa_2 \sigma_1 \sigma_2] \nonumber \\ 
   \cos \theta_{s-} e^{i\theta_{s+}} &=& i \kappa_3 \sigma_3 \mu_0 [i
    \kappa_1 \sigma_1 \mu_2 +\mu_1 \kappa_2 \sigma_2] 
  \end{eqnarray}
Noting that $g_1^{(2)} \to -\infty$ in Eq.~(\ref{eq:majorana-mass-0})
implies that $\mu_0$ develops long range order for $\langle
\theta_{c,A}\rangle=0$, we find that in the low energy limit the
expressions in (\ref{eq:ising-su2-2}) reduce to:
\begin{eqnarray}
  \cos \phi_{s-} e^{i\phi_{s+}} &\sim&\mu_3  [\mu_1 \mu_2 + \kappa_1
    \kappa_2 \sigma_1 \sigma_2], \nonumber \\
\cos \theta_{s-} e^{i\theta_{s+}} &\sim& i \kappa_3 \sigma_3 [i
    \kappa_1 \sigma_1 \mu_2 +\mu_1 \kappa_2 \sigma_2] 
\end{eqnarray}

Introducing the Pauli spin matrices $\tau_c=-\frac i 2 \epsilon_{abc}\kappa_b
\kappa_c$  these expressions are easily seen to reduce to the expression of the
spin 1/2 primary fields of the $SU(2)_2$ Wess-Zumino-Novikov-Witten
model\cite{novikov82_wz,witten_wz,polyakov83_wz,zamolodchikov_fateev,allen} in terms of Ising
fields. Moreover, the expression of the spin currents
(\ref{eq:majorana-currents}) also reduces to the $SU(2)_2$ form.  

Thus the theory describing the spin excitations  at low energy reduces to a 
 $SU(2)_1$ WZNW model (that describes the spin excitations of the band
 0) coupled with a $SU(2)_2$ WZNW model (that describes the spin
 excitations of the bands $\pm 1$)  by a term: 
\begin{equation}\label{eq:coupling}
 \lambda\int dx ~ \mathrm{tr}
( g_1 \mathbf{\sigma})(x) \cdot \mathrm{tr}( g_2 \mathbf{\sigma})(x), 
\end{equation}
\noindent and a marginal current-current interaction term. Power
counting shows that the term (\ref{eq:coupling}) is relevant with RG
dimension $5/4$. Therefore, it is reasonable to treat first this
relevant term, as was done in the case of two-leg spin
ladders.\cite{shelton_spin_ladders} 
For analyzing the effect of the interaction~(\ref{eq:coupling}) on the
spin spectrum, is convenient to introduce 
a coset representation\cite{goddard_cosets}:
$SU(2)_1\times SU(2)_2 \sim SU(2)_3 \times \text{TIM}$ where
$\text{TIM}$ stands for  
the tricritical Ising model\footnote{This coset was used in an
unpublished work of P. Lecheminant on three-leg spin ladders.}.  
With the coset decomposition, we can rewrite the  WZNW fields as:
\begin{eqnarray}
  \label{eq:coset-fields}
  g_1&=&\epsilon_{TIM} g_3 \nonumber \\ 
  g_2&=&\sigma_{TIM} g_3, 
\end{eqnarray}
where $g_k$ is the spin 1/2 $ SU(2)_k$ field, $\epsilon_{TIM}$ is
the energy operator of the Tricritical Ising Model of dimension 
$\frac 1 {10}$, 
$\sigma_{TIM}$ is the spin operator of the tricritical 
Ising model of dimension $\frac 3 {80}$. Using the Operator Product 
Expansion of the
$\text{TIM}$ $\epsilon_{TIM} \sigma_{TIM} \sim (\sigma
+\sigma')_{TIM}$ from Ref.~\onlinecite{difrancesco97_book} (p. 224) 
we can rewrite the interaction (\ref{eq:coupling}) as:
\begin{eqnarray}
  \label{eq:coset-interaction}
  \lambda'\int dx  \sigma_{TIM} \mathrm{tr}( g_3 \mathbf{\sigma})(x)
\cdot \mathrm{tr}( g_3 \mathbf{\sigma})(x),  
\end{eqnarray}
where only the most relevant term has been kept. The interaction is
now brought to the form of a self-coupling for the $SU(2)_3$ WZNW
model. Now, we simplify this
self coupling by using a second coset\cite{zamolodchikov85_parafermions}
 representation,
$SU(2)_3 \sim U(1) \times Z_3$ where $Z_3$ represents the critical
3-state Potts model (or equivalently the 3-state clock
model)\cite{jose_planar_2d}  
and $U(1)$ represents a free bosonic field described
by the Hamiltonian: 
\begin{eqnarray}\label{eq:hamiltonian-u1} 
  H=\int \frac {dx}{2\pi}\left[v_F K (\pi \Pi)^2 +
  \frac{v_F}{K}(\partial_x \phi)^2\right],  
\end{eqnarray}
\noindent where $K=1$. This coset representation was used in
Ref.~\onlinecite{cabra_spin_s} in a study of the Haldane gap 
in spin-S chains, and from
now on our treatment follows this work closely. 
 We write the
components of the fundamental field as $g^{[m,m']}_3$ with $m,m'=\pm
1/2$. We have from
Refs.~\onlinecite{zamolodchikov_fateev,zamolodchikov85_parafermions} the  relations: 
\begin{eqnarray}
  g_3^{[\frac 1 2,\frac 1 2]} =e^{-i \sqrt{\frac 2 3} \theta} \sigma_1, \\
  g_3^{[-\frac 1 2,\frac 1 2]} =e^{-i \sqrt{\frac 2 3} \phi} \mu_1,
\end{eqnarray}
\noindent where $\sigma_1,\mu_1$ are the order and disorder parameters
  of  the 3-state clock model. 
With this, we can  rewrite the interaction as:
\begin{multline}
\lambda^{''} \int dx  \sigma_{TIM} 
\biggl(\sigma_1 (\sigma_1)^\dagger + \mu_1 \mu_1^\dagger/2 \\
 + e^{-2i \sqrt{\frac 2 3} \phi} \mu_1^2 
 + e^{-2i \sqrt{\frac 2 3} \phi} (\mu_1^\dagger)^2 \biggr)
\end{multline}
Then we use the properties of the 3-state clock
model\cite{zamolodchikov85_parafermions}: 
$\sigma_1^\dagger=\sigma_2$ $\sigma_1^2=\sigma_2$ and similarly with
$\mu \leftrightarrow \sigma$, together with the Operator 
Product Expansion $\sigma_1 \sigma_2 \sim \epsilon$. This allows to
reduce the above interaction term to: 
\begin{equation}
\lambda^{''} \int dx  \sigma_{TIM} \biggl(
\epsilon_{Z_3}  + e^{-2i \sqrt{\frac 2 3} \phi} \mu_2 
+ e^{-2i \sqrt{\frac 2 3} \phi} \mu_1 \biggr).   
\end{equation}
Now let us make the assumption that the TIM 
develops a long range order and so
does the 3-state clock model. Let us assume further that the 3-state 
clock model is in the low temperature phase,  with 
$\mu_{1,2}$ disordered. Only the bosonic field $\phi$ can \emph{a
  priori} remain gapless. In order to determine whether $\phi$ indeed
remains gapless, we have to consider the perturbations generated by
the disordered operators $\mu_{1,2}$. It is straightforward to see 
that these terms yield a perturbation: 
\begin{eqnarray}\label{eq:perturbation}
  \lambda_0 \cos 2 \sqrt{6} \phi
\end{eqnarray}
for the $U(1)$ theory. The operators of the $SU(2)_3$ theory then 
reduce to\cite{cabra_spin_s}:
\begin{align}\label{eq:uv-ir-spins}
  n^+&=  tr (g\sigma^+) \sim e^{-i \sqrt{\frac 2 3} \theta}\\  
 n^z&=  tr (g\sigma^z) \sim (e^{-i \sqrt{\frac 2 3} \phi} \mu_1+ e^{i
\sqrt{\frac 2 3} \phi} \mu_2)\\ \nonumber  
 & \quad \times ( e^{-2i \sqrt{\frac 2 3} \phi} \mu_2 + e^{-2i
\sqrt{\frac 2 3} \phi} \mu_1) \sim   e^{-i \sqrt{6} \phi} +  e^{i
\sqrt{6} \phi} 
\end{align} 
The $SU(2)$ symmetry of the system imposes that $n^+$ and $n^z$ have
the same scaling dimension. Therefore, at this new fixed point, one
must have $K=1/3$. Hence, after a rescaling $\phi =\tilde{\phi}/\sqrt{3}$ 
and $\theta=\sqrt{3}\tilde{\theta}$, $\tilde{K}=3K$ 
the expressions (\ref{eq:hamiltonian-u1}) and (\ref{eq:uv-ir-spins}) 
reduce to the ones of
the $SU(2)_1$ case\cite{giamarchi_book_1d}, with a
perturbation~(\ref{eq:perturbation}) which is  
marginal. Two regimes are possible depending on whether $\lambda_0$ is
marginally relevant
or marginally irrelevant. 
In the first case, a spin gap is obtained. In the second case,
no spin gap is obtained and the system has the same spin correlation
as a free  $SU(2)_1$ model up to logarithmic corrections.\cite{solyom_revue_1d,giamarchi_logs,affleck_su2_logs} 
In order to predict which phase is realized, we have to consider the
flow of the Luttinger exponent of $\phi$. If the fixed point is
approached from the side where the
perturbation~(\ref{eq:perturbation}) is irrelevant, then we can expect
the fixed point to be stable. 
Since at the origin the Luttinger exponent is $K=1$, and
at the fixed point it is $K=1/3$, the flow is indeed on the
side where (\ref{eq:perturbation}) is marginally irrelevant. Thus, we find
gapless spin modes at the fixed point with both triplet 
and singlet superconductivity. However, the 
logarithmic corrections induced by the marginally irrelevant perturbation at the 
$SU(2)_1$ fixed point are known\cite{solyom_revue_1d,giamarchi_logs,affleck_su2_logs}
to lead to dominant triplet superconductivity fluctuations. Triplet
superconductivity is thus naturally expected in the present case. In
this respect, we note that a similar situation arises in single chain
system where the renormalization group predicts dominant triplet
superconducting fluctuations in the vicinity of a spin density wave
phase.\cite{solyom_revue_1d} Since in Refs.~\onlinecite{Tang:2001,Takesue:2005}
the superconductivity appears to be sensitive to the application of a
magnetic field, it is likely that the intertube coupling tends to
better stabilize the singlet superconductivity with respect to the
triplet one.  
If we assume that $\mu_1$ is ordered, then we find that $\phi$ is also long 
range ordered. As a result, $n^+$ is short range ordered 
while $n^z$ or $\epsilon=tr(g_3)$ is long range ordered. Since the system has to
be rotationally symmetric, the only solution is to have $\epsilon$
long range ordered and $n^z$ short range ordered. In that case, the
system has a spin gap, and only the singlet superconducting  order
parameters exhibits quasi-long range order.


\section{Conclusion}
\label{sec:conclusion}

 In conclusion, by means of fermionic renormalization, abelian and
 non-abelian bosonization we have analyzed  the low 
energy properties of a three band one dimensional model deduced from
the band structure of cylindrical small radius (5,0) nanotubes.  
 We have found that this system possess a specific instability,
besides the usual single band and two band models
instabilities. This instability corresponds to the development of
superconducting fluctuations in the nanotube. Within our approach, in
 the absence of a spin gap, 
triplet superconductivity fluctuations are expected to be dominant du
 to logarithmic corrections, with
subdominant  singlet superconductivity fluctuations. 

   This new instability is tightly related to the symmetry of our
three band model, and more precisely to the new couplings $u$ and
$v$.  In our model, in the presence of these couplings, 
the spin excitations are either fully
gapped leaving only a C1S0 phase as in the two-leg ladder\cite{varma_2chain,penc_2chain,fabrizio_2ch_rg,nagaosa_2ch,khveshenko_2chain,schulz_2chains,nagaosa_chiral_anomaly_1d,balents_2ch,shelton_tj_ladder,schulz_moriond} or they 
are described by a $SU(2)_1$ WZNW model leading to a C1S1 phase 
as in a single chain Hubbard model. This is in contrast to previous
studies of three legs model with different symmetries, which included
only 2 bands couplings as opposed to the 3 bands couplings  $u$
or $v$ : in these models,  a C2S1
phase was
found\cite{arrigoni_3chain,kimura96_3chain,lin97_nchains}. Technically,
this difference lies in the ordering of the field $\theta_{c,B}$,
directly related to the presence of the $v$ coupling. Note that the 3
bands nature of the $v$ coupling also induced the technical problem of
the Klein factor discussed in this paper.  

 Let us finally relate our results to previous studies on the (5,0)
nanotubes. In Ref.~\onlinecite{Kamide:2003}, only a subset of the
couplings of the present model was considered, which did not include
the $u$ and $v$ term. Hence this new superconducting instability was
not discussed. In Ref.~\onlinecite{Gonzalez:2005}, Gonzalez and
Perfetto studied the same model as ours, by means of a
renormalization group procedure. The nature of the phase was determined {\it
via} the scaling behavior of correlations functions, as opposed to
the bosonization procedure used in this paper. In
Ref.~\onlinecite{Gonzalez:2005}, it was found that the dominant
instability would be a charge density wave coupling the bands $\pm 1$,
with subdominant spin density wave  fluctuations, whereas we find that 
charge density wave fluctuations are suppressed.  
The origin of this discrepancy is that in
Ref.~\onlinecite{Gonzalez:2005}, only specific initial conditions
were considered, with  initial values of some
couplings so large that they render  a one-loop
renormalization group approach questionable.  It appears likely that 
the initial conditions chosen in Ref.~\onlinecite{Gonzalez:2005}
strongly favor a two band instability between the band $\pm
1$. Indeed, a divergent $g_1^{(1)}$ as we found in
Sec.~\ref{sec:single-band} indeed leads to a reinforcement of charge 
density wave fluctuations between the bands $\pm 1$.

Along these lines, let us mention that it is difficult to determine
which of the possible mechanisms, including the one proposed in this
paper, actually takes place in a (5,0) nanotube. Indeed, as opposed to
theoretical approaches of conventional larger nanotubes, our
one-dimensional electronic model is using the band structure provided
by  ab-initio calculations as an input.
Since these methods already include a renormalization of the band structure by
a fraction of the electronic interactions, an estimate of the bare
coupling in our one dimensional model based on an unrenormalized
Coulomb interaction as in Ref.~\onlinecite{Egger:1997} is likely to
lead to misleading results by overestimating the effect of some
interactions.  As a result, we can only propose a classification of
the various fixed points at weak coupling and characterize the
possible scenarii, with the usual hypothesis in one-dimensional
systems that the weak coupling behavior and the strong coupling
behavior are continuously connected.\cite{schulz_houches_revue} A related remark is that the gaps
calculated in any weak coupling approximation are generally very small.
However, in the real system, where interaction strength can be
expected to be comparable to the bandwidth, since Luttinger exponents are found in the range\cite{bockrath_review} $0.2-0.5$,  the real gaps can be much
higher than those estimated in a weak coupling treatment. Thus, the
present treatment cannot lead to a realistic estimate of critical
temperatures.   

 Another aspect of the physics to consider is the possibility of a 
pseudo-Peierls transitions in this small radius
nanotubes\cite{Connetable:2005}. Indeed, the approach of this paper
is based on the band structure of numerical approach which did not
consider the possibility of a cylindrical geometry breaking. If such
a phenomena was to happen, an indication of strong electron-phonon coupling in the system, other  mechanisms for superconductivity could
 occur, but their description is beyond the scope of the
present paper. Finally, let us mention that the experimental results
on the superconducting transition in these small
nanotubes\cite{Tang:2001,Takesue:2005} suggest that a real three
dimensional superconducting phase transition takes place. A complete
understanding of these results must also include a  coupling between 
the nanotubes to stabilize the superconducting fluctuations at nonzero
temperature. However, since the gap in the zeolite matrix is of order
4eV, an intertube Josephson coupling term would be  {\it a priori} strongly
suppressed by the presence of the insulator between the tubes.

 We thank X. Blase for very stimulating discussions which initiated
this work. P. Pujol is also acknowledge for insightful remarks on the
non-abelian bosonization approach of section \ref{sec:NonAbelian}.

\appendix

\section{Derivation of the RG equations}
\label{app:RG-deriv}
 To express the RG equations in a simpler form, we first define 
 rescaled couplings as 
 \begin{align}
& \tilde{g}_{i}^{(j)} = \frac{1}{2\pi v_{F}} g_{i}^{(j)} 
\quad ; \quad
\tilde{f}^{(i)} =  \frac{1}{2\pi v_{F}}f^{(i)}
\\
& \tilde{u}  =  \frac{1}{2\pi v_{F}}u 
\quad ; \quad
\tilde{v}  = \frac{1}{2\pi v_{F}} v 
\end{align}
The scaling equations read in terms of this couplings 
\begin{subequations}
\label{eq:RGeqs}
\begin{align}
\partial_{l} \tilde{g}^{(1)} = &    
-2 (\tilde{g}^{(1)})^2  - 4 \tilde{u} \tilde{v}  \\
\partial_{l} \tilde{g}^{(2)} = &    
- (\tilde{g}^{(1)})^2 - 2 \tilde{u}^2 - 2  \tilde{v}^2 \\
\partial_{l} \tilde{g}_{1}^{(1)} = & 
- 2 (\tilde{g}_{1}^{(1)})^2
-2 \tilde{g}_{1}^{(2)} \tilde{g}_{2}^{(1)} - 2 \tilde{u} \tilde{v} \\
\partial_{l} \tilde{g}_{1}^{(2)} = &
- 2 \tilde{g}_{1}^{(1)} \tilde{g}_{2}^{(1)}
- 2 \tilde{g}_{1}^{(2)} \tilde{g}_{2}^{(2)} 
+ 2 \tilde{g}_{1}^{(2)} \tilde{g}_{4}^{(2)}
- \tilde{u}^2 - \tilde{v}^2  \\
\nonumber \partial_{l} \tilde{g}_{2}^{(1)} = &
 - 2 \tilde{g}_{1}^{(1)} \tilde{g}_{1}^{(2)}
 - 2 \tilde{g}_{2}^{(1)} \tilde{g}_{2}^{(2)}
 + 2 \tilde{g}_{1}^{(2)} \tilde{g}_{4}^{(1)} \\
 & -4 \tilde{g}_{2}^{(1)} \tilde{g}_{4}^{(1)}
 + 2 \tilde{g}_{2}^{(1)} \tilde{g}_{4}^{(2)} - 2  \tilde{u} \tilde{v}  \\
\partial_{l} \tilde{g}_{2}^{(2)} = &
 -  (\tilde{g}_{1}^{(1)})^2 -  (\tilde{g}_{1}^{(2)})^2
 -  (\tilde{g}_{2}^{(1)})^2 -   \tilde{u}^2 -  \tilde{v}^2   \\
\partial_{l} \tilde{g}_{4}^{(1)} = & 
+ 2 \tilde{g}_{1}^{(2)} \tilde{g}_{2}^{(1)}
-2 ( \tilde{g}_{2}^{(1)} )^2
-2  ( \tilde{g}_{4}^{(1)} )^2 \\
\partial_{l} \tilde{g}_{4}^{(2)} = &  
(\tilde{g}_{1}^{(2)})^2
-   (\tilde{g}_{4}^{(1)})^2    \\
\partial_{l} \tilde{f}^{(1)} = &     
-2 (\tilde{f}^{(1)})^2 + 2 \tilde{u} \tilde{v} -2 \tilde{v}^2   \\
\partial_{l} \tilde{f}^{(2)} = &    
-  (\tilde{f}^{(1)})^2 +  \tilde{u}^2  \\ 
\nonumber \partial_{l} \tilde{u} = &
 \left( 2 \tilde{f}^{(2)} -   \tilde{g}_{1}^{(2)} 
-  \tilde{g}^{(2)} -  \tilde{g}_{2}^{(2)} \right) \tilde{u} \\
 & - \left(  \tilde{g}^{(1)} + \tilde{g}_{1}^{(1)} 
+ \tilde{g}_{2}^{(1)} \right) \tilde{v}
  \\ \nonumber
\partial_{l} \tilde{v} =&
-  \left( -2 \tilde{f}^{(1)} +  \tilde{g}^{(1)} 
+  \tilde{g}_{1}^{(1)} +  \tilde{g}_{2}^{(1)}\right) \tilde{u} \\
& - \left( 4 \tilde{f}^{(1)} - 2 \tilde{f}^{(2)} 
+  \tilde{g}_{1}^{(2)} +  \tilde{g}^{(2)} +  \tilde{g}_{2}^{(2)}\right) \tilde{v}
\end{align}
\end{subequations}

\section{Bosonization}
\label{app:bosonization}

In this appendix, we provide some technical details on our
bosonization approach. We first give the bosonized expressions for
the three interband operators $f^{(1)},g_{1}^{(2)}$ and $v$ within
the usual ``condensed matter'' convention. These expressions show why
this common convention is not suitable for the present analysis. 
 Second, we provide the detailed expressions of the various 
order parameters  of our three band model within the field
theoretical convention used and defined in the text. We use the
definition and conventions of the section \ref{sec:phases} of the
paper. 

\subsection{Derivation of the bosonized form of interband 
interactions}

We consider the interactions $f^{(1)},g_{1}^{(2)}$ 
that involve a transfer of fermions between two different bands. 
These terms can be read from the action (\ref{eq:formal-Sint}). We
use the condensed matter convention of section
\ref{sec:CondensedMatter}. 
\begin{widetext} 
The $g_1^{(2)}$ term can be expressed in the form :
\begin{equation}\label{eq:g12-condmat}
-\frac{g_1^{(2)}} {(2\pi a)^2} \left[ e^{i 2 \phi_{c,A}} \sum_\sigma \left[
   e^{i 2\sigma \phi_{s,-}}\eta_{R,1,\sigma} \eta_{L,1,\sigma}
 \eta_{L,-1,\sigma} \eta_{R,-1,\sigma}
+ e^{i 2\sigma \theta_{s,-}} \eta_{R,1,\sigma} \eta_{L,1,-\sigma}
    \eta_{L,-1,-\sigma} \eta_{R,-1,\sigma} \right] 
+ \text{H. c.} \right]
\end{equation}
The contribution of $f^{(1)}$ is composed of a term quadratic in the
$\Phi$ fields and a term:
\begin{align}\label{eq:f1-condmat}
-\frac{ f^{(1)}} {(2\pi a)^2} \sum_\sigma \biggl\{
& e^{-i \sigma \sqrt{2} (\theta_{s,0} -\phi_{s,0})} \eta_{R,0,\sigma}
\eta_{R,0,-\sigma} e^{i\sigma(\theta_{s,+}+\phi_{s,+})} 
\nonumber \\
&\qquad  \times \left[
e^{i\sigma(\theta_{s,-}+\phi_{s,-})} \eta_{L,1,-\sigma}
\eta_{L,1,\sigma} + e^{-i\sigma(\theta_{s,-}+\phi_{s,-})}
\eta_{L,-1,-\sigma} \eta_{L,-1,\sigma} \right] 
\nonumber \\
 + & e^{-i \sigma \sqrt{2} (\theta_{s,0} + \phi_{s,0})}
\eta_{L,0,\sigma} \eta_{L,0,-\sigma}
e^{i\sigma(\theta_{s,+}-\phi_{s,+})} 
\nonumber \\
& \qquad \times \left[
e^{i\sigma(\theta_{s,-}-\phi_{s,-})} \eta_{R,1,-\sigma}
\eta_{R,1,\sigma} + e^{-i\sigma(\theta_{s,-}-\phi_{s,-})}
\eta_{R,-1,-\sigma} \eta_{R,-1,\sigma}\right]
\biggr\}
\end{align}
Note that in nonabelian
bosonization\cite{witten_wz,knizhnik_wz,zamolodchikov_fateev,affleck_houches}
the term $f^{(1)}$ is seen as the interaction of a $SU(2)_1$
current associated with the spin excitations of the band $0$ with a
$SU(2)_2$ current composed of the sum of the $SU(2)_1$ currents
associated with the spin excitations of the bands $\pm 1$.
And finally, the contribution of $v$ reads:
\begin{align}\label{eq:v-condmat}
\frac {v}{(2\pi a)^2} \biggl\{ 
e^{-i \sqrt{3} \theta_{c,B}} e^{i \phi_{c,A}} \biggl[
& e^{i\sigma(\sqrt{2} \theta_{s,0} - \theta_{s,+}+ \phi_{s,-})} 
\eta_{R,1,\sigma}  \eta_{L,-1,\sigma} 
\eta_{R,0,\sigma}  \eta_{L,0,\sigma}
\nonumber \\
+& e^{i\sigma (\sqrt{2} \phi_{s,0} + \phi_{s,+}-\theta_{s,-})}
\eta_{R,1,\sigma}  \eta_{L,-1,-\sigma}  
\eta_{R,0,-\sigma} \eta_{L,0,\sigma} \biggr]   
\nonumber \\ 
+  e^{-i \sqrt{3} \theta_{c,B}} e^{-i \phi_{c,A}}  \biggl[
& e^{i\sigma (\sqrt{2} \theta_{s,0} - \theta_{s,+} -\phi_{s,-}) }
  \eta_{L,1,\sigma}  \eta_{R,-1,\sigma}  
  \eta_{L,0,\sigma}  \eta_{R,0,\sigma} 
\nonumber \\
+ & e^{i\sigma (\sqrt{2} \phi_{s,0} + \phi_{s,+} + i\theta_{s,-})}
 \eta_{L,1,\sigma}  \eta_{R,-1,-\sigma} 
 \eta_{L,0,-\sigma} \eta_{R,0,\sigma}  \biggr] 
+\text{H. c.} \biggr\}  
\end{align}
\end{widetext}

\subsection{Order parameters}
\label{sec:opera-boso}

In this last section, we switch back to the field theoretical
convention for bosonization, defined in the section
\ref{sec:FieldTheory}. 

\subsubsection{Superconductivity}

We consider the following order parameters  the formation of
singlet superconductivity in the nanotube :  
\begin{eqnarray}
  \label{eq:op-supra}
  O_0(x)&=&\sum_\sigma \psi_{R,0,\sigma} \psi_{L,0,-\sigma} \\ 
  O_1(x)&=&\sum_\sigma \psi_{R,1,\sigma} \psi_{L,-1,-\sigma} \\
  O_{-1}(x)&=&\sum_\sigma \psi_{R,-1,\sigma} \psi_{L,1,-\sigma} 
\end{eqnarray}

Using the bosonization decomposition introduced in the text, we can
express these order parameters for superconductivity  into bosonized
variables : 
\begin{subequations}\label{eq:superconductivity}
\begin{align}
O_0(x)&= \frac{i}{\pi\alpha}
e^{-i\sqrt{2}\left[\frac{\theta_{c+}}{\sqrt{3}} -\sqrt{\frac 2 3}
\theta_{c,B}\right]}
 \sin  \sqrt{2}\phi_{s,0}
\\
O_1(x)&= \frac{i}{\pi\alpha} 
e^{i\sqrt{2}\left(
  \frac{\theta_{c+}}{\sqrt{3}}
+ \frac{\theta_{c,A}}{\sqrt{2}}
+ \frac{\theta_{c,B}}{\sqrt{6}}
\right)}
\sin (\phi_{s,+}+\phi_{s,-})
\\
O_1(x)&= \frac{i}{\pi\alpha} 
e^{i\sqrt{2}\left(
\frac{\theta_{c+}}{\sqrt{3}}
- \frac{\theta_{c,A}}{\sqrt{2}}
+ \frac{\theta_{c,B}}{\sqrt{6}}
\right)}
\sin (\phi_{s,+}-\phi_{s,-})
\end{align}
\end{subequations}

 Note that we have not considered the triplet superconductivity order
parameters. Indeed, they naturally possess the same charge part than
the singlet superconductivity operators, and only differ by their
spin part. Since the spin part is more conveniently treated within the
non-abelian bosonization formalism (see Sec.~\ref{sec:NonAbelian}), it
is not necessary to give an explicit expression of the triplet
operators here, since thay can be obtained from the expression of the
operators~(\ref{eq:superconductivity}) in nonabelian bosonization by
the substitution $tr(g) \to tr(g\mathbf{\sigma})$.

\subsubsection{Charge density waves}

Besides superconductivity, one can also expect to observe charge
density wave order to develop at low temperature in a quasi-1D
system. 
The various charge density wave order operators, labelled by their
ordering momentum vector, are defined as : 
\begin{align}\label{eq:op-cdw}
O_{(2k_{F_0},0)}(x)&=\sum_\sigma \psi^\dagger_{R,0,\sigma} \psi_{L,0,\sigma}\\ 
O_{(2k_{F_1},0)}(x)&=\sum_\sigma \psi^\dagger_{R,1,\sigma} \psi_{L,1,\sigma}\\ 
O_{(2k_{F_{-1}},0)}(x)&=
    \sum_\sigma \psi^\dagger_{R,-1,\sigma} \psi_{L,-1,\sigma}\\ 
O_{(k_{F_0}-k_{F_1},K_y)}(x)&=
    \sum_\sigma \psi^\dagger_{R,0,\sigma} \psi_{L,-1,\sigma}\\ 
O_{(k_{F_0}-k_{F_1},-K_y)}(x)&=
    \sum_\sigma \psi^\dagger_{R,0,\sigma} \psi_{L,1,\sigma}\\ 
O_{(2k_{F_1},2K_y}(x)&= \sum_\sigma \psi^\dagger_{R,1,\sigma} \psi_{L,+1,\sigma}\\ 
O_{(2k_{F_1},-2K_y}(x)&= \sum_\sigma \psi^\dagger_{R,1,\sigma} \psi_{L,-1,\sigma}
\end{align}
 We have considered only the charge density wave  operators of the
form $\psi^\dagger_{R,m,\sigma} \psi_{L,m',\sigma}$, as those of the
form $\psi^\dagger_{R,m,\sigma} \psi_{R,m',\sigma}$ are current that
cannot develop quasi-long range order. 

The corresponding bosonized expressions of these charge density wave order
parameters is easily obtained and read 
\begin{widetext}
\begin{align}\label{eq:CDW1}
O_{(2k_{F_0},0)}(x) &= 
\frac{-i}{\pi\alpha}
e^{i\sqrt{2}\left[\frac{\phi_{c+}}{\sqrt{3}} 
                 -\frac{2 \phi_{c,A}}{\sqrt{6}}\right]} 
\cos \sqrt{2} \phi_{s,0}
\\ 
O_{(2k_{F_1},0)}(x) &=
\frac 1 {2\pi\alpha}
e^{i\left[\frac{2 \phi_{c+}}{\sqrt{3}}
         - \theta_{c,A}
         +\sqrt{\frac{2}{3}}\phi_{c,B}\right]}  
\sum_\sigma  e^{i\sigma(\phi_{s,+}-\theta_{s,-})} 
\eta_{1,\sigma} \eta_{-1,\sigma} 
\\  
O_{(2k_{F_{-1}},0)}(x) &= 
\frac 1 {2\pi\alpha}
e^{i\left[\frac{2 \phi_{c+}}{\sqrt{3}}
         + \theta_{c,A}
         +\sqrt{\frac{2}{3}}\phi_{c,B}\right]}  
\sum_\sigma  e^{i\sigma(\phi_{s,+}+\theta_{s,-})} 
\eta_{-1,\sigma} \eta_{1,\sigma}
\\ 
O_{(-k_{F_0}-k_{F_{1}},K_y)}(x)&= 
\frac 1 {2\pi\alpha}
e^{i\left[ \frac{2\phi_{c+}}{\sqrt{6}}
          -\frac{1}{\sqrt{2}} (\theta_{cA}+\phi_{cA})
          +\frac{3}{\sqrt{6}  (\theta_{cB}-\phi_{cB})}
\right]}
 \sum_\sigma 
e^{i\sigma \left[ 
 -\frac 1 2 (\theta_{s,+}+\theta_{s,-}-\phi_{s,+}-\phi_{s,-})
 -\frac{\theta_{s,0}-\phi_{s,0}}{\sqrt{2}} 
\right]} 
\eta_{0,\sigma} \eta_{-1,\sigma}  
\\
O_{(-k_{F_0}-k_{F_{1}},K_y)}(x)&= 
\frac 1 {2\pi\alpha}
e^{i\left[ \frac{2\phi_{c+}}{\sqrt{6}}
          +\frac{1}{\sqrt{2}} (\theta_{cA}+\phi_{cA})
          +\frac{3}{\sqrt{6}  (\theta_{cB}-\phi_{cB})}
\right]}
 \sum_\sigma 
e^{i\sigma \left[ 
 -\frac 1 2 (\theta_{s,+}-\theta_{s,-}-\phi_{s,+}+\phi_{s,-})
 -\frac{\theta_{s,0}-\phi_{s,0}}{\sqrt{2}} 
\right]} 
\eta_{0,\sigma} \eta_{1,\sigma}  
\\
O_{(2k_{F_{1}},2K_y)} & = 
\frac{i}{2\pi\alpha}
e^{i\sqrt{2}\left[\frac{\phi_{c+}}{\sqrt{3}}
                 +\frac{\phi_{cA}}{\sqrt{2}}
                 +\frac{\phi_{cB}}{\sqrt{6}}
 \right]}
\sum_\sigma
e^{i\sigma(\phi_{s,+}+\phi_{s,-})} 
\\ \label{eq:CDW2}
O_{(2k_{F_{1}},- 2K_y)} & = 
\frac{i}{2\pi\alpha}
e^{i\sqrt{2}\left[\frac{\phi_{c+}}{\sqrt{3}}
                 -\frac{\phi_{cA}}{\sqrt{2}}
                 +\frac{\phi_{cB}}{\sqrt{6}}
 \right]}
\sum_\sigma
e^{i\sigma(\phi_{s,+}-\phi_{s,-})} 
\end{align}
\end{widetext}

\section{Equivalence of Bosonization Conventions}
\label{app:klein}

In this appendix, we discuss a general case of 
 1D fermions with $N$ ``flavors'' since the results are of more 
general applicability than the nanotube with three bands at the Fermi
level that we have considered in this paper.  
 The bosonized representation of fermion operators used in 
condensed matter literature amounts to
write~\cite{haldane_bosonisation,schulz_moriond,delft_bosonization,%
senechal_bosonization_revue}: 
\begin{eqnarray}
  \label{eq:bosonize-1}
\psi_{R,n}&=&\frac 1 {\sqrt{2\pi\alpha}} e^{-i\Phi_{R,n}(x)} \eta_{R,n}, \\
 \psi_{L,n}&=&\frac 1 {\sqrt{2\pi\alpha}} e^{i\Phi_{L,n}(x)} \eta_{L,n},
\end{eqnarray}

where:
\begin{eqnarray}
\lbrack\Phi_{R,n}(x),\Phi_{R,n'}(x')\rbrack&=&i\pi\delta_{n,n'} 
\mathrm{sign}(x-x'), \\  
\lbrack\Phi_{R,n}(x),\Phi_{R,n'}(x')\rbrack&=&-i\pi \delta_{n,n'} 
\mathrm{sign}(x-x'), \\
\lbrack\Phi_{R,n}(x),\Phi_{L,n'}(x')\rbrack&=&0,
\end{eqnarray}  
and the Majorana fermion operators $\eta_{\nu,n}$ satisfy:
\begin{eqnarray}
\label{eq:majorana-1}
\{\eta_{\nu,n},\eta_{\nu',n'}\}=2\delta_{\nu,nu'}\delta_{n,n'} .
\end{eqnarray}
In the field theoretical literature, an apparently different 
bosonized representation is used\cite{halpern:1975,zuber_77,ha:1984}: 
\begin{eqnarray}
  \label{eq:bosonize-2}
\psi_{R,n}&=&\frac 1 {\sqrt{2\pi\alpha}} 
e^{-i\tilde{\Phi}_{R,n}(x)} \eta_{n}, \\
\psi_{L,n}&=&\frac 1 {\sqrt{2\pi\alpha}} 
e^{i\tilde{\Phi}_{L,n}(x)} \eta_{n},
\end{eqnarray}
where this time:  
\begin{eqnarray}
\lbrack \tilde{\Phi}_{R,n}(x),\tilde{\Phi}_{R,n'}(x')\rbrack=
i\pi \delta_{n,n'}\mathrm{sign}(x-x'),  \\
\lbrack\tilde{\Phi}_{L,n}(x),\tilde{\Phi}_{L,n'}(x')\rbrack=
-i\pi \delta_{n,n'} \mathrm{sign}(x-x'), \\
\lbrack\tilde{\Phi}_{R,n}(x),\tilde{\Phi}_{L,n'}(x')\rbrack=
i\pi \delta_{n,n'} 
\end{eqnarray}

and the Majorana fermion operators $\eta_{n}$ satisfy:
\begin{eqnarray}
  \label{eq:majorana-2}
  \{\eta_{n},\eta_{n'}\}=2\delta_{n,n'}.
\end{eqnarray}
While in the condensed-matter 
bosonization for a given flavor 
the right and left bosonic fields are made commuting and
there is one Majorana fermion associated with the left mover and another 
Majorana fermion associated with the right mover, in the field theoretical
representation, there is only one Majorana fermion for each flavor. The price 
to pay for this is to make the commutator of the chiral fields non-zero. 
An application of this representation in condensed matter physics is the
the derivation of the bosonized form of the doubled 
Ising model\cite{zuber_77,schroer_ising,boyanovsky_ising}.  

To show that these two representations are in fact equivalent, let
us introduce the conjugate variables $Q_n$ and $P_n$ such that: 
\begin{eqnarray}
  \lbrack Q_n,\tilde{\Phi}_{\nu,n'}(x) \rbrack &=&0, \\
  \lbrack P_n,\tilde{\Phi}_{\nu,n'}(x) \rbrack &=&0, \\
  \lbrack Q_n,P_m \rbrack & =&i\pi \delta_{n,m}, 
\end{eqnarray}
and let us write: 
\begin{eqnarray}
  \bar{\Phi}_{R,n}&=&\tilde{\Phi}_{R,n}-\sqrt{\frac \pi 2} (Q_n -P_n), \\
   \bar{\Phi}_{L,n}&=&\tilde{\Phi}_{L,n}+\sqrt{\frac \pi 2} (Q_n +P_n).
\end{eqnarray}
We have: 
$[\bar{\Phi}_{R,n},\bar{\Phi}_{L,n}]=i\pi -\pi([Q_n,P_n]-[P_n,Q_n])/2=0$.
Then the fermion operators are rewritten as: 
\begin{eqnarray}
  \label{eq:bosonize-3}
 \psi_{R,n}&=&\frac 1 {\sqrt{2\pi\alpha}} 
e^{-i\bar{\Phi}_{R,n}(x)} e^{i\sqrt{\frac \pi 2} (Q_n -P_n)} \eta_{n}, \\ 
  \psi_{L,n}&=&\frac 1 {\sqrt{2\pi\alpha}} 
e^{i\bar{\Phi}_{L,n}(x)}  e^{i\sqrt{\frac \pi 2} (Q_n + P_n)} \eta_{n},
\end{eqnarray}
and one has:
\begin{eqnarray}
&&   e^{i\sqrt{\frac \pi 2} (Q_n -P_n)} \eta_{n}  
     e^{i\sqrt{\frac \pi 2} (Q_n + P_n)} \eta_{n} \nonumber \\ 
&=&  e^{i\sqrt{\frac \pi 2} (Q_n + P_n)} \eta_{n} 
      e^{i\sqrt{\frac \pi 2} (Q_n -P_n)} \eta_{n} 
      e^{-\frac \pi 2 [Q_n + P_n, Q_n -P_n]}, \nonumber \\
&=& - e^{i\sqrt{\frac \pi 2} (Q_n + P_n)} \eta_{n} 
      e^{i\sqrt{\frac \pi 2} (Q_n -P_n)} \eta_{n}.  
\end{eqnarray}
Therefore, we can define a set of Majorana fermion operators: 
\begin{eqnarray}
\eta_{R,n}&=& e^{i\sqrt{\frac \pi 2} (Q_n -P_n)} \eta_{n}, \\
\eta_{L,n}&=& e^{i\sqrt{\frac \pi 2} (Q_n + P_n)} \eta_{n}
\end{eqnarray}
which satisfy the commutation relations (\ref{eq:majorana-1}). 
As a result, the
representation (\ref{eq:bosonize-3}) and 
the representation (\ref{eq:bosonize-2}) are  equivalent to 
the representation (\ref{eq:bosonize-1}).

It is well known that one can also define bosonization using
non-chiral fields\cite{giamarchi_book_1d} $\theta_n,\phi_n$ given by: 
\begin{eqnarray}\label{nonchiral}
 \phi_n = \frac 1 2 (\Phi_{L,n}+\Phi_{R,n}) \\ 
 \theta_n= \frac 1 2 (\Phi_{L,n}-\Phi_{R,n}) 
\end{eqnarray}

In the case of the bosonization procedure (\ref{eq:bosonize-1}), the non-chiral
fields have the commutation relation:
\begin{eqnarray}\label{commutator1}
\lbrack \phi_n(x),\phi_m(x') \rbrack &=& 0, \\
\lbrack \theta(x)_n,\theta_m(x') \rbrack &=& 0, \\ 
\lbrack \theta(x)_n,\phi_m(x') \rbrack &=& 
-i\frac \pi 2 \delta_{n,m} \mathrm{sign} (x-x'),
\end{eqnarray}

In the case of the bosonization procedure (\ref{eq:bosonize-2}), the
non-chiral fields can be defined similarly but they have the
commutation relation: $ \lbrack \tilde{\theta(x)}_n,\tilde{\phi}_m(x')
\rbrack = -i \pi \delta_{n,m} \Theta (x-x')$, where $\Theta$ is the
Heaviside function.  The relation between the two sets of chiral
fields is:
\begin{eqnarray}\label{nonchiral-map}
  \tilde{\phi}_n(x)&=&\phi_n(x) +\sqrt{\frac \pi 2} Q_n \\
  \tilde{\theta}_n(x)&=&\theta_n(x)+\sqrt{\frac \pi 2} P_n
\end{eqnarray}

An interpretation of $Q_n$ is that it is proportional to 
an operator counting the number 
of fermions (both right and left moving) of type $n$. 
$P_n$ is the proportional to 
the phase conjugate to this fermion number, and thus
must be compactified. 

Note that if we perform a rotation on the fields $\theta_n$ and
$\phi_n$, this rotation preserves the commutation
relations~(\ref{commutator1}). This is a crucial property.  It also
preserves the commutation relations in the case of field theoretic
bosonization, and obviously it preserves the relation of commutation
of the zero modes $P$ and $Q$. Therefore, one always goes from the
condensed matter convention to the field theoretical convention by
using the same shift of the non-chiral fields~(\ref{nonchiral-map}) by
zero modes.

When converting products of two fermion operators from the condensed matter 
convention to the field-theoretical convention, the following rules apply: 
$\eta_{R/L,n} \eta_{R/L,n'} \to \eta_{n} \eta_{n'}$ for different 
species $n\ne n'$ and for a given specie $n$ :
\begin{eqnarray}
e^{-i \Phi_{R,n}} \eta_{R,n}  e^{i \Phi_{L,n}} \eta_{L,n} 
\to i e^{i ( \tilde{\Phi}_{L,n}-\tilde{\Phi}_{R,n})} = 
i e^{i 2\tilde{\phi}_n},    \nonumber \\  
e^{i \Phi_{R,n}} \eta_{R,n}  e^{i \Phi_{L,n}} \eta_{L,n} 
\to -i e^{i ( \Phi_{L,n}-\Phi_{R,n})}=-i e^{i 2\tilde{\theta}_n},    
\nonumber \\ 
\end{eqnarray}
We close this section with two remarks. First, the field theoretical
representation can be understood naturally on a semi-infinite system
extending to $+\infty$, with an open boundary condition at the
origin\cite{fabrizio_open_electron_gas} : the origin is then sent to
$-\infty$ to recover an infinite system. The non-zero value of the
commutator of the right and left moving field possess then the physical
interpretation of left movers being reflected into right movers at
$-\infty$. As a second remark let us note that when applying
the condensed matter
bosonization to the double Ising model, one finds that the operator
$\cos \phi$ does not have the same commutation with the the fermion
field as the products of Ising disorder fields\cite{schroer_ising},
whereas this relation is satisfied in the field theoretical
bosonization.  The reason for this difference 
can be inferred from our first
remark :  the construction of the product of disorder fields as a
string is made for a system with an open boundary condition at the
origin.  This boundary condition is included from the start in field
theoretical bosonization but not in the condensed matter
bosonization. Thus, when using condensed matter bosonization, Klein
factors must appear in the expression of the products of the Ising
disorder fields as a function of $\cos \phi$.


\begin{thebibliography}{10}

\bibitem{Kociak:2001}
M. Kociak {\it et~al.}, Phys. Rev. Lett. {\bf 86},  2416  (2001).

\bibitem{Tang:2001}
Z. Tang {\it et~al.}, Science {\bf 292},  2462  (2001).

\bibitem{Takesue:2005}
I.Takesue {\it et~al.}, cond-mat/0509466, to be published in Phys. Rev. Lett.
  (unpublished).

\bibitem{Blase:1994}
X. Blase, L.~X. Benedict, E.~L. Shirley, and S.~G. Louie, Phys. Rev. Lett. {\bf
  72},  1878  (1994).

\bibitem{Sedeki:2002}
A. S{\'e}d{\'e}ki, L.~G. Caron, and C. Bourbonnais, Phys. Rev. B {\bf 65},
  140515(R)  (2002).

\bibitem{Connetable:2005}
D. Conn{\'e}table, G.-M. Rignanese, J.-C. Charlier, and X. Blase, Phys. Rev.
  Lett. {\bf 94},  015503  (2005).

\bibitem{Barnett:2005}
R. Barnett, E. Demler, and E. Kaxiras, Solid State Communications {\bf 135},
  335  (2005).

\bibitem{Kamide:2003}
K. Kamide, T. Kimura, M. Nishida, and S. Kurihara, Phys. Rev. B {\bf 68},
  024506  (2003).

\bibitem{Gonzalez:2005}
J. Gonz{\'a}lez and E. Perfetto, Phys. Rev. B {\bf 72},  205406  (2005).

\bibitem{Egger:1997}
R. Egger and A.~O. Gogolin, Phys. Rev. Lett. {\bf 79},  5082  (1997).

\bibitem{egger_nanotubes}
R. Egger and A. Gogolin, Eur. Phys. J. B {\bf 3},  281  (1998).

\bibitem{balents97_nanotubes}
L. Balents and M.~P.~A. Fisher, Phys. Rev. B {\bf 55},  R11973  (1997).

\bibitem{kane97_nanotubes}
C. Kane, L. Balents, and M.~P.~A. Fisher, Phys. Rev. Lett. {\bf 79},  5086
  (1997).

\bibitem{yoshioka99_nanotubes}
H. Yoshioka and A.~A. Odintsov, Phys. Rev. Lett. {\bf 82},  374  (1999).

\bibitem{odintsov99_nanotubes}
A.~A. Odintsov and H. Yoshioka, Phys. Rev. B {\bf 59},  R10457  (1999).

\bibitem{odintsov_nanotubes}
A.~A. Odintsov and H. Yoshioka,  in {\em Low-Dimensional Systems: Interactions
  and transport properties}, Vol.~544 of {\em Lecture Notes in Physics}, edited
  by T. Brandes (Springer, Heidelberg, 2000), p.\ 97.

\bibitem{lin98_nanotubes}
H.-H. Lin, Phys. Rev. B {\bf 58},  4963  (1998).

\bibitem{levitov_nanotube}
L.~S. Levitov and A.~M. Tsvelik, Phys. Rev. Lett. {\bf 90},  016401  (2003).

\bibitem{nersesyan03_nanotube}
A.~A. Nersesyan and A.~M. Tsvelik, Phys. Rev. B {\bf 68},  235419  (2003),
  cond-mat/0305311.

\bibitem{li01_small_radius_nanotubes}
Z.~M. Li {\it et~al.}, Phys. Rev. Lett. {\bf 87},  127401  (2001).

\bibitem{machon02_nanotube_bands}
M. Mach\'on {\it et~al.}, Phys. Rev. B {\bf 66},  155410  (2002).

\bibitem{liu02_nanotube_bands}
H.~J. Liu and C.~T. Chan, Phys. Rev. B {\bf 66},  115416  (2002).

\bibitem{miyake03_nanotubes_gw}
T. Miyake and S. Saito, Phys. Rev. B {\bf 155424},  155424  (2003).

\bibitem{arrigoni_3chain}
E. Arrigoni, Phys. Lett. A {\bf 215},  91  (1996), cond-mat/9509145.

\bibitem{kimura96_3chain}
T. Kimura, K. Kuroki, and H. Aoki, Phys. Rev. B {\bf 54},  R9608  (1996).

\bibitem{schulz_moriond}
H.~J. Schulz,  in {\em Correlated fermions and transport in mesoscopic
  systems}, edited by T. Martin, G. Montambaux, and J. {Tran Thanh Van}
  (Editions fronti\`eres, Gif sur Yvette, France, 1996), p.\ 81.

\bibitem{lin97_nchains}
H.-H. Lin, L. Balents, and M.~P.~A. Fisher, Phys. Rev. B {\bf 56},  6569
  (1997).

\bibitem{rice97_3chain}
T.~M. Rice, S. Haas, M. Sigrist, and F.-C. Zhang, Phys. Rev. B {\bf 56},  14655
   (1997).

\bibitem{white98_3chains_dmrg}
S.~R. White and D.~J. Scalapino, Phys. Rev. B {\bf 57},  3031  (1998).

\bibitem{kimura98_3chain}
T. Kimura, K. Kuroki, and H. Aoki, J. Phys. Soc. Jpn. {\bf 67},  1377  (1998).

\bibitem{yonemitsu99_3chains_dmrg}
K. Yonemitsu, J. Low Temp. Phys {\bf 117},  1765  (1999).

\bibitem{ledermann00_nchains}
U. Ledermann, K.~L. Hur, and T.~M. Rice, Phys. Rev. B {\bf 62},  16383  (2000).

\bibitem{tsuchiizu01_3chains}
M. Tsuchiizu and Y. Suzumura, J. Phys. Chem. Solids {\bf 62},  427  (2001).

\bibitem{deMartino:2004}
A.~D. Martino and R. Egger, Phys. Rev. B {\bf 70},  014508  (2004).

\bibitem{witten_wz}
E. Witten, Commun . Math. Phys. {\bf 92},  455  (1984).

\bibitem{knizhnik_wz}
V.~G. Knizhnik and A.~B. Zamolodchikov, Nucl. Phys. B {\bf 247},  83  (1984).

\bibitem{novikov82_wz}
S.~P. Novikov, Usp. Fiz. Nauk {\bf 37},  3  (1982).

\bibitem{polyakov83_wz}
A.~M. Polyakov and P.~B. Wiegmann, Phys. Lett. B {\bf 131},  121  (1983).

\bibitem{affleck_wz}
I. Affleck, Nucl. Phys. B {\bf 265},  409  (1986).

\bibitem{affleck_strongcoupl}
I. Affleck and F.~D.~M. Haldane, Phys. Rev. B {\bf 36},  5291  (1987).

\bibitem{solyom_revue_1d}
J. S{\'o}lyom, Adv. Phys. {\bf 28},  209  (1979).

\bibitem{schulz_houches_revue}
H.~J. Schulz,  in {\em Mesoscopic quantum physics, Les Houches LXI}, edited by
  E. Akkermans, G. Montambaux, J.~L. Pichard, and J. Zinn-Justin (Elsevier,
  Amsterdam, 1995), p.\ 533.

\bibitem{Krotov:1997}
Y.~A. Krotov, D.-H. Lee, and S.~G. Louie, Phys. Rev. Lett. {\bf 78},  4245
  (1997).

\bibitem{varma_2chain}
C.~M. Varma and A. Zawadowski, Phys. Rev. B {\bf 32},  7399  (1985).

\bibitem{penc_2chain}
K. Penc and J. S{\'o}lyom, Phys. Rev. B {\bf 41},  704  (1990).

\bibitem{fabrizio_2ch_rg}
M. Fabrizio, Phys. Rev. B {\bf 48},  15838  (1993).

\bibitem{khveshenko_2chain}
D.~V. Khveshenko and T.~M. Rice, Phys. Rev. B {\bf 50},  252  (1994).

\bibitem{nagaosa_2ch}
N. Nagaosa, Sol. State Comm. {\bf 94},  495  (1995).

\bibitem{schulz_2chains}
H.~J. Schulz, Phys. Rev. B {\bf 53},  R2959  (1996).

\bibitem{nagaosa_chiral_anomaly_1d}
N. Nagaosa and M. Oshikawa, J. Phys. Soc. Jpn. {\bf 65},  2241  (1996).

\bibitem{balents_2ch}
L. Balents and M.~P.~A. Fisher, Phys. Rev. B {\bf 53},  12133  (1996).

\bibitem{yoshioka_2ch}
H. Yoshioka and Y. Suzumura, Phys. Rev. B {\bf 54},  9328  (1996).

\bibitem{shelton_tj_ladder}
D.~G. Shelton and A.~M. Tsvelik, Phys. Rev. B {\bf 53},  14036  (1996).

\bibitem{lee99_2ch}
Y.-W. Lee and Y.-L. Lee, Phys. Rev. B {\bf 59},  1290  (1999).

\bibitem{kuroki94_2ch}
K. Kuroki and H. Aoki, Phys. Rev. Lett. {\bf 72},  2947  (1994).

\bibitem{abramovici05_2ch}
G. Abramovici, J.~C. Nickel, and M. {H\'eritier}, Phys. Rev. B {\bf 72},
  045120  (2005).

\bibitem{ando05_nanotube_review}
T. Ando, J. Phys. Soc. Jpn. {\bf 74},  777  (2005).

\bibitem{cardy_scaling}
J.~L. Cardy, {\em Scaling and Renormalization in Statistical Physics}, {\em
  Cambridge Lecture Notes in Physics} (Cambridge University Press, Cambdridge,
  UK, 1996).

\bibitem{lin00_scaling}
H.-H. Lin, cond-mat/0010011 (unpublished).

\bibitem{konik02_symmetry}
R.~M. Konik, H. Saleur, and A.~W.~W. Ludwig, Phys. Rev. B {\bf 66},  075105
  (2002).

\bibitem{giamarchi_book_1d}
T. Giamarchi, {\em Quantum Physics in one Dimension}, Vol.~121 of {\em
  International series of monographs on physics} (Oxford University Press,
  Oxford, UK, 2004).

\bibitem{gogolin_book}
A.~O. Gogolin, A.~A. Nersesyan, and A.~M. Tsvelik, {\em Bosonization and
  Strongly Correlated Systems} (Cambridge University Press, Cambridge, 1999).

\bibitem{luther_exact}
A. Luther and V.~J. Emery, Phys. Rev. Lett. {\bf 33},  589  (1974).

\bibitem{emery_revue_1d}
V.~J. Emery,  in {\em Highly Conducting One-Dimensional Solids}, edited by
  J.~T. Devreese, R.~P. Evrard, and V.~E. van Doren (Plenum Press, New York and
  London, 1979).

\bibitem{haldane_bosonisation}
F.~D.~M. Haldane, J. Phys. C {\bf 14},  2585  (1981).

\bibitem{heidenreich_bosonisation}
R. Heidenreich, R. Seiler, and D.~A. Uhlenbrock, J. Stat. Phys. {\bf 22},  27
  (1980).

\bibitem{delft_bosonization}
J. {von Delft} and H. Schoeller, Ann. Phys. (Leipzig) {\bf 7},  225  (1998).

\bibitem{jose_planar_2d}
J.~V. Jos{\'e}, L.~P. Kadanoff, S. Kirkpatrick, and D.~R. Nelson, Phys. Rev. B
  {\bf 16},  1217  (1977).

\bibitem{lecheminant02_sdsg}
P. Lecheminant, A.~O. Gogolin, and A.~A. Nersesyan, Nucl. Phys. B {\bf 639},
  502  (2002).

\bibitem{finkelstein_2ch}
A.~M. Finkelstein and A.~I. Larkin, Phys. Rev. B {\bf 47},  10461  (1993).

\bibitem{zamolodchikov_fateev}
A.~B. Zamolodchikov and V.~A. Fateev, Yad. Fiz. {\bf 43},  1031  (1986), [Sov.
  J. Nucl. Phys. {\bf 43} p. 657 (1986)].

\bibitem{shelton_spin_ladders}
D.~G. Shelton, A.~A. Nersesyan, and A.~M. Tsvelik, Phys. Rev. B {\bf 53},  8521
   (1996).

\bibitem{allen}
D. Allen and D. S{\'e}n{\'e}chal, Phys. Rev. B {\bf 55},  299  (1997).

\bibitem{emery_2channel}
V.~J. Emery and S.~A. Kivelson, Phys. Rev. B {\bf 46},  10812  (1992).

\bibitem{kadanoff_gaussian_model}
L.~P. Kadanoff, Ann. Phys. (NY) {\bf 120},  39  (1979).

\bibitem{zuber_77}
J.~B. Zuber and C. Itzykson, Phys. Rev. D {\bf 15},  2875  (1977).

\bibitem{schroer_ising}
B. Schroer and T.~T. Truong, Nucl. Phys. B {\bf 144},  80  (1978).

\bibitem{ogilvie_ising}
M. Ogilvie, Ann. Phys. (NY) {\bf 136},  273  (1981).

\bibitem{boyanovsky_ising}
D. Boyanovsky, Phys. Rev. B {\bf 39},  6744  (1989).

\bibitem{nersesyan01_ising_review}
A.~A. Nersesyan,  in {\em New Theoretical approaches to strongly correlated
  systems}, Vol.~23 of {\em Nato Science Series II: Mathematics, Physics and
  Chemistry}, edited by A.~M. Tsvelik (Kluwer Academic Publishers, Dordrecht,
  2001), p.\ 89.

\bibitem{goddard_cosets}
P. Goddard, A. Kent, and D. Olive, Phys. Lett. B {\bf 152},  88  (1985).

\bibitem{difrancesco97_book}
P. {Di~Francesco}, P. Mathieu, and D. {S\'en\'chal}, {\em Conformal Field
  Theory} (Springer, Heidelberg, 1997).

\bibitem{zamolodchikov85_parafermions}
A.~B. Zamolodchikov and V.~A.~. Fateev, Zh. {\'E}ksp. Teor. Fiz. {\bf 89},  380
   (1985), [Sov. Phys. JETP \textbf{62} 215 (1986)].

\bibitem{cabra_spin_s}
D.~C. Cabra, P. Pujol, and C. {von Reichenbach}, Phys. Rev. B {\bf 58},  65
  (1998).

\bibitem{giamarchi_logs}
T. Giamarchi and H.~J. Schulz, Phys. Rev. B {\bf 39},  4620  (1989).

\bibitem{affleck_su2_logs}
I. Affleck, D. Gepner, H.~J. Schulz, and T. Ziman, J. Phys. A {\bf 22},  511
  (1989).

\bibitem{bockrath_review}
M. Bockrath {\it et~al.}, Nature {\bf 397},  598  (1999).

\bibitem{affleck_houches}
I. Affleck,  in {\em Fields, Strings and Critical Phenomena}, edited by E.
  Brezin and J. Zinn-Justin (Elsevier Science Publishers, Amsterdam, 1988).

\bibitem{senechal_bosonization_revue}
D. {S\'en\'echal},  in {\em Theoretical methods for strongly correlated
  electrons}, {\em CRM series in mathematical physics}, edited by D.
  S\'en\'chal, A.-M. Tremblay, and C. Bourbonnais (Springer, Heidelberg, 2004),
  p.\ 139.

\bibitem{halpern:1975}
M.~B. Halpern, Phys. Rev. D {\bf 12},  1684  (1975).

\bibitem{ha:1984}
Y.~K. Ha, Phys. Rev. D {\bf 29},  1744  (1984).

\bibitem{fabrizio_open_electron_gas}
M. Fabrizio and A.~O. Gogolin, Phys. Rev. B {\bf 51},  17827  (1995).

\end{thebibliography}
\end{document}